\documentclass[aps, prb, twocolumn, superscriptaddress]{revtex4-2}

\usepackage{amssymb}
\usepackage{amsmath}
\usepackage{amsfonts}
\usepackage{bm}
\usepackage{mathtools}

\usepackage{color}
\usepackage{xcolor}

\usepackage{graphicx}
\usepackage{grffile}

\usepackage{array}

\usepackage{hyperref}
\hypersetup{colorlinks=true, allcolors=blue}
\usepackage{url}

\newcommand{\addressASNR}{ASNR/PSN-RES/SEMIA/LSMA Centre d’études de Cadarache, Saint-Paul-lez-Durance, 13115, France}
\newcommand{\addressMpiSusMat}{Max Planck Institute for Sustainable Materials, Düsseldorf, 40327, Germany}
\newcommand{\addressCINaM}{Aix-Marseille Université, CNRS, CINaM UMR 7325, Marseille 13288, France}
\newcommand{\addressUniMich}{Department of Mechanical Engineering, University of Michigan, Ann Arbor, Michigan 48109, USA}

\begin{document}
    \title{Thermodynamics of complex crystals from a generalized mean-field bond model}
    
    \author{Raynol Dsouza}
    \email{r.dsouza@mpi-susmat.de}
    \affiliation{\addressASNR}
    \affiliation{\addressMpiSusMat}
    
    \author{Thomas D. Swinburne}
    \email{tswin@umich.edu}
    \affiliation{\addressUniMich}
    \affiliation{\addressCINaM}
    
    \author{J\"org Neugebauer}
    \affiliation{\addressMpiSusMat}
    
    \begin{abstract}
    We extend our previously introduced mean-field bond model from unary fcc lattices to crystals with multiple chemical species and sublattices, accurately predicting finite-temperature thermodynamics without statistical sampling.
    Our generalization reformulates the model in terms of the Wyckoff positions of the crystal, representing each by one atom, and solving for the bond density of the reference bond of each of its neighbor shells.
    Bond-bond correlation functions are built from the zero-temperature force constants, allowing the deviation of every bond in the crystal to be expressed through linear functions of the reference bond of a shell.
    Thermodynamic self-consistency is enforced by requiring a single virial temperature for the whole crystal, and compatibility of average bond lengths with crystal volume.
    Model predictions are validated against molecular dynamics simulations of unary, binary, and ternary crystals using Morse-type bonding potentials, parametrized for each shell from the force constants of empirical interatomic potentials at several isotropic strains.
    In each case, the predicted bond densities and thermodynamics agree with the simulations to within their statistical uncertainty.
    We discuss strategies for combining the model with bonding potentials parametrized from first principles and foundation models to predict the finite-temperature thermodynamics of complex crystals at \emph{ab initio} accuracy without statistical sampling.
    \end{abstract}
    \date{November 24, 2026}
    \maketitle


    \section{Introduction}
    \label{sec:introduction}

    Many technologically relevant properties of materials, such as thermal expansion, heat capacity, and phase stability, vary with temperature, in large part through lattice vibrations~\cite{Wallace1972}.
    In contrast to zero-temperature properties, for which a single relaxed configuration is sufficient, these properties are ensemble averages, conventionally evaluated over a large number of thermally sampled configurations~\cite{Frenkel2002}.
    Even with significant advances in modern high-performance computing, evaluating such averages with density-functional theory (DFT)~\cite{Hohenberg1964, Kohn1965} remains prohibitively expensive, and finite-temperature thermodynamic properties are commonly computed analytically within the harmonic or quasiharmonic approximation (QHA)~\cite{Baroni2001, Togo2023}.

    Although the QHA captures the implicit anharmonicity due to the volume dependence of phonon frequencies, it neglects the explicit anharmonicity due to the underlying potential energy surface at each volume, the omission of which has been shown to lead to qualitatively incorrect thermodynamic trends at elevated temperatures~\cite{Grabowski2009, Glensk2014, Glensk2015}.
    In principle, explicit anharmonicity can be computed exactly using thermodynamic integration (TI), which integrates the ensemble-averaged energy difference between the target and a reference system along a suitable coupling path~\cite{Kirkwood1935, Frenkel1984}.
    In practice, however, every point along this path requires a converged ensemble average, and the number of sampled configurations needed grows with the difference between the potential energy surfaces of the two systems.
    Accelerating this convergence has motivated the use of intermediate reference systems that are cheaper to sample yet closer to the target potential energy surface, whose free energies can be upsampled to \emph{ab initio} accuracy using free energy perturbation~\cite{Zwanzig1954} by running DFT calculations on a small subset of their configurations~\cite{Grabowski2009, Duff2015, Grabowski2019, Jung2023}.
    Alternatively, effective-harmonic methods, such as the temperature-dependent effective potential method~\cite{Hellman2011, Hellman2013, Shulumba2017, Knoop2024} and the stochastic self-consistent harmonic approximation~\cite{Errea2014, Bianco2017, Monacelli2021}, extract temperature-dependent force constants from forces on thermally displaced configurations and approximate thermodynamic properties using temperature-renormalized phonon frequencies.
    While these methods avoid the coupling path of TI, they still require sampling at each volume and temperature of interest.
    
    Since every sampled configuration requires an evaluation of the forces on each atom, the computational cost of going beyond the QHA is determined by the number of such evaluations and the associated cost of each one.
    Machine-learning interatomic potentials (MLIPs) reduce the cost of each force evaluation while approaching the accuracy of the DFT data on which they are trained~\cite{Behler2007, Bartok2010, Drautz2019}. 
    In contrast to the cubic scaling of DFT, MLIPs scale linearly with the number of atoms and offer a compromise between the efficiency of empirical potentials~\cite{Daw1984, Mishin2005} and the accuracy of DFT.
    MLIPs have since also evolved into universal potentials or foundation models~\cite{Batzner2022, Kavanagh2026, Chen2022, Batatia2022, Batatia2025, Pozdnyakov2023, Mazitov2025, Bochkarev2024, Lysogorskiy2026} that are trained on databases spanning most of the periodic table~\cite{Deng2023, BarrosLuque2026}.
    However, their accuracy relative to DFT for thermodynamic and vibrational properties has been benchmarked predominantly using quantities derived from the zero-temperature potential energy surface, namely formation energies and thermodynamic stability~\cite{Riebesell2025}, harmonic phonons~\cite{Loew2025}, and lattice thermal conductivities computed from second- and third-order force constants~\cite{Pota2026}.
    Assessing their accuracy for finite-temperature thermodynamics requires sampling for each crystal, volume, and temperature of interest, with the cost per force evaluation set by the architecture of the foundation model.
    Since this cost cannot be reduced without changing the model itself, the total computational effort can only be lowered by reducing the number of force evaluations.

    An alternative that eliminates finite-temperature force evaluations altogether is the mean-field bond model, proposed by Swinburne \textit{et al.}~\cite{Swinburne2020}.
    It builds on the observation of Glensk \textit{et al.}~\cite{Glensk2015} that the explicit anharmonic free energy of fcc crystals is reproduced to within a few meV/atom of \emph{ab initio} TI by a lattice of anharmonic first-nearest-neighbor bonds.
    Swinburne \textit{et al.} showed that in the isochoric ensemble, the finite-temperature thermodynamics of such a bond lattice can be evaluated without sampling, by deriving the bond density analytically from a polarized bonding potential through a mean-field treatment of the geometric constraints between bonds and enforcing thermodynamic self-consistency.
    In our previous work~\cite{Dsouza2024}, we extended the mean-field bond model to bonding potentials of arbitrary functional form, by parametrizing them from a few static displacements of an atom in a target potential following the local anharmonic approximation~\cite{Glensk2015}.
    We further refined it with an energy-volume correction to account for the interactions in the target potential not captured by the bonding potentials, thereby enabling its use in the isobaric ensemble.
    Applied to specialized MLIPs for fcc Al and Cu, the extended model reproduced the anharmonic free energy to within a few meV/atom of that of TI up to the melting temperature, with a similar accuracy for thermal expansion.
    Both formulations, however, only account for first-nearest-neighbor contributions to the bond density, and are therefore restricted to crystals with a single atom per primitive cell.
    The former suffices for fcc crystals, where anharmonicity is dominated by the first-nearest-neighbor bonds~\cite{Glensk2015}, but not for lattices such as bcc, where the next-nearest-neighbor bonds contribute comparably~\cite{Dsouza2025thesis}.
    The latter restriction excludes all ordered multicomponent crystals, as well as multisublattice crystals with a single chemical species, such as hcp.

    In this work, we generalize the model to arbitrary neighbor shells in crystals with multiple sublattices and chemical species by reformulating it in terms of the Wyckoff positions of the crystal.
    We solve for the bond density of the reference bond of each shell of one representative atom per Wyckoff position, and obtain the thermodynamic properties of the crystal as atomic-fraction-weighted averages over the Wyckoff positions.
    The generalization retains the sampling-free character of the original model, with the correlations between bonds derived from the harmonic response of the crystal lattice rather than imposed by its geometry alone.
    The coupling between bonds requires only the force constants of the relaxed crystal, and is independent of temperature and atomic mass.
    Thermodynamic self-consistency is enforced through volume constraints for each shell and a single virial temperature constraint for the crystal as a whole, with the bonds coupled to the reference bond of each shell taken around both of its ends, such that bonds shared between two Wyckoff positions are described identically from either end.

\begin{figure*}
        \includegraphics[width=\textwidth]{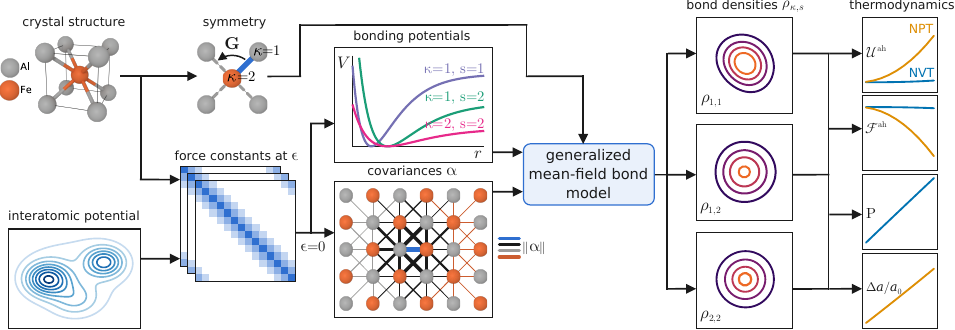}
        \caption{Overview of the approach presented in this work, illustrated for B2 AlFe. The crystal structure and its symmetry, which generates every bond of a shell from its reference bond through $\mathbf G$, reduce the crystal to one representative atom per Wyckoff position $\kappa$ and one reference bond per shell $s$. The force constants of the interatomic potential at several isotropic strains are used to parametrize the Morse-type bonding potentials, while the force constants at $\epsilon=0$ are used to obtain the bond-bond covariances on bond-centered clusters, from which the self-consistent model yields the bond densities $\rho_{\kappa,s}$ and the thermodynamic properties, namely the anharmonic internal energy $\mathcal U^{\mathrm{ah}}$ and the anharmonic free energy $\mathcal F^{\mathrm{ah}}$ in the NVT and NPT ensembles, the pressure $P$ (NVT), and the lattice expansion $\Delta a/a_0$ (NPT).}
        \label{fig:workflow}
    \end{figure*}

    Since the bonding potentials can be parametrized from any source, for example, from static displacements of the representative atom of a Wyckoff position in a target potential~\cite{Glensk2015, Dsouza2024}, the generalized mean-field bond model can be applied directly to either DFT or a foundation model.
    In order to separate the error of the model from that of the bonding potentials themselves, we use identical Morse-type bonding potentials, parametrized for each shell from the force constants of empirical interatomic potentials at several isotropic strains, in both the model and the molecular dynamics (MD) simulations used for validating the model.
    The comparison therefore assesses the generalized model alone, and not how well the bonding potentials reproduce the empirical interatomic potentials from which they were parametrized.
    An overview of the approach presented in this work is given in Fig.~\ref{fig:workflow}.

    This paper is organized as follows:
    Sec.~\ref{sec:theory} develops the theory of the generalized model, and Sec.~\ref{sec:methodology} describes the choice of crystal structures, the parametrization of the bonding potentials, and the computational details of the model and the MD simulations.
    Sec.~\ref{sec:results} compares the results of the model with those from MD for the chosen crystal structures.
    Sec.~\ref{sec:discussion} discusses the computational cost and limitations of the model, as well as strategies for its application to DFT and foundation models. Sec.~\ref{sec:conclusion} presents the conclusions.

    \section{Theory}
    \label{sec:theory}
    
        \subsection{Symmetry equivalence of atoms and bonds}
        \label{sec:theory-symmetry}
    
        Consider a crystal with $N$ atoms in $d$ dimensions confined to a volume $V_0(1+\epsilon)^d$ under an isotropic strain $\epsilon$, such that $\epsilon = 0$ corresponds to the relaxed crystal at zero temperature.
        For crystals whose cell remains cubic at any volume and temperature, $\epsilon$ describes the deformation of the cell exactly, whereas for those with lower symmetry, it fixes the shape of the cell at its zero-temperature value.
        An atom $i$ at position $\mathbf{R}_i$ belongs to one of $K$ symmetry-distinct Wyckoff positions $\kappa(i) \in \{1, \dots, K\}$, determined from the space group of the relaxed crystal~\cite{Togo2024}.
        Each Wyckoff position $\kappa$ is occupied by $N_\kappa$ atoms of a single chemical species with atomic fraction $c_\kappa \equiv N_\kappa/N$.
        Since atoms belonging to the same Wyckoff position are related by space-group symmetry operations and have identical local environments, we select an arbitrary atom $i_\kappa$ to represent each $\kappa$.

        The neighbors of atom $i_\kappa$ can be grouped into $S_\kappa$ shells based on their distance from it, with each shell $s$ containing $m_{\kappa,s}$ neighbors indexed by $l \in \{1, \dots, m_{\kappa,s}\}$.
        In cases where neighbors at the same distance do not form a single orbit under the site-symmetry group of atom $i_\kappa$, we treat each additional orbit as a distinct shell.
        We then define the $l$-th neighbor bond in shell $s$ of Wyckoff position $\kappa$ as
        \begin{equation}
            \mathbf b_{i_\kappa}^{(s,l)} \equiv \mathbf{R}_{n_{s,l}(i_\kappa)} - \mathbf{R}_{i_\kappa} \in \mathbb{R}^d,
            \label{eq:bondvec}
        \end{equation}
        where $n_{s,l}(i_\kappa)$ is the index of the $l$-th neighbor in shell $s$ of atom $i_\kappa$, and write $\mathbf b_\kappa^{(s,l)} \equiv \mathbf b_{i_\kappa}^{(s,l)}$.

        In the absence of lattice vibrations, the instantaneous bond $\mathbf{b}_\kappa^{(s,l)}$ reduces to its equilibrium value $\mathbf{a}^{\epsilon}_{\kappa,s,l}$ under the applied strain $\epsilon$.
        We take the $l=1$ bond of shell $s$ as its reference bond $\mathbf b_{\kappa,s} \equiv \mathbf b_\kappa^{(s,1)}$, with equilibrium value $\mathbf{a}^{\epsilon}_{\kappa,s} \equiv \mathbf{a}^{\epsilon}_{\kappa,s,1}$, from which all other equilibrium bonds in $s$ follow as
        \begin{equation}
            \mathbf{a}^{\epsilon}_{\kappa,s,l} = \mathbf{G}_{\kappa,s,l}\,\mathbf{a}^{\epsilon}_{\kappa,s},
            \label{eq:orbitgen}
        \end{equation}
        where $\mathbf{G}_{\kappa,s,l}$ is the site-symmetry operation of atom $i_\kappa$ that maps the reference bond onto the $l$-th bond, with $\mathbf{G}_{\kappa,s,1} = \mathbb I$.
        The matrices $\mathbf{G}_{\kappa,s,l}$ generalize the symmetry operations $\mathbf{G}_{1l}$ of Ref.~\cite{Swinburne2020}, defined for a single shell and a single Wyckoff position, to arbitrary shells and Wyckoff positions in the crystal.

        Since all neighbor bonds in a shell are symmetry-equivalent, a single bonding potential $V_{\kappa,s}$ is sufficient to characterize the interactions associated with a $(\kappa,s)$ pair.
        Additionally, a bond connecting atom $i_\kappa$ to a target atom $i_{\kappa'}$ is the same physical bond when viewed from either position, and must therefore be described by the same bonding potential from both ends.
        We refer to the two shells that contain the same physical bonds as a reciprocal pair $(\kappa,s)\leftrightarrow(\kappa',s')$, whose two shells coincide when the bond joins two atoms of the same Wyckoff position, and the site symmetry of $i_\kappa$ maps $\mathbf a^\epsilon_{\kappa,s}$ onto $-\mathbf a^\epsilon_{\kappa,s}$.
        The reference bonds from each end of the reciprocal pair are related by $\mathbf a^\epsilon_{\kappa',s'} = -\mathbf Q_{\kappa,s}\,\mathbf a^\epsilon_{\kappa,s}$, where $\mathbf Q_{\kappa,s}$ is the rotational part of a space-group operation, with their respective bonding potentials satisfying
        \begin{equation}
            V_{\kappa',s'}\!\left(-\mathbf Q_{\kappa,s}\mathbf b\right) = V_{\kappa,s}(\mathbf b).
            \label{eq:reciprocity}
        \end{equation}
        We note that this reciprocity is a consistency requirement imposed on the set $\{V_{\kappa,s}\}$, rather than a property that follows automatically from assigning an independent bonding potential to each $(\kappa,s)$ pair.
        Counting every physical bond once from each of its ends further relates the bond counts of the pair as $c_\kappa m_{\kappa,s} = c_{\kappa'} m_{\kappa',s'}$.
        
        Taken together, the above relations reduce the crystal from $N$ atoms to $K$ representative atoms, each fully characterized by a single reference bond and a single bonding potential per shell, from which every other bond in the crystal and its associated interaction follows by symmetry.
        This symmetry reduction of the bonds is exact and follows from the space group of the relaxed crystal alone.

        \subsection{The generalized bond lattice}
        \label{sec:theory-bondlattice}
    
        Although the interactions in the crystal are fully specified by the $K$ representative atoms and their associated bonding potentials, the total energy is a property of the crystal as a whole, and must therefore account for the contribution of every atom in the crystal.
        Since $V_{\kappa,s}$ is defined relative to the reference bond $\mathbf{b}^{\epsilon}_{\kappa,s}$, evaluating the contribution of the $l$-th neighbor bond in shell $s$ requires first rotating that bond to the reference frame using $\mathbf{G}_{i,s,l}^{-1}$.
        The total energy of the crystal, grouped by Wyckoff position $\kappa$, is then given by
        \begin{equation}
            U(\{\mathbf{b}\}) = \frac{1}{2} \sum_{\kappa=1}^{K} \sum_{i=1}^{N_\kappa}
                \sum_{s=1}^{S_\kappa} \sum_{l=1}^{m_{\kappa,s}}
                V_{\kappa,s}\!\left(\mathbf{G}_{i,s,l}^{-1}\mathbf{b}_i^{(s,l)}\right),
            \label{eq:energy}
        \end{equation}
        where $i$ indexes the $N_\kappa$ atoms belonging to $\kappa$, and the factor $1/2$ corrects the double-counting of each bond when summing over both ends.

        For a general atom $i$, the full set of neighbor bonds $\{\mathbf b\}$ is overcomplete, and must be restricted to the $3N$-dimensional constraint surface $C(\{\mathbf b\})=0$ defined by material compatibility~\cite{Swinburne2020}.
        Using pairwise antisymmetry, compatibility requires that any two bonds $\mathbf b^{(s,l)}$ and $\mathbf b^{(s',l')}$ originating from a common atom $i$ satisfy
        \begin{equation}
            \mathbf b_i^{(s,l)} - \mathbf b_{n_{s',l'}(i)}^{(s,l)} = \mathbf b_i^{(s',l')} - \mathbf b_{n_{s,l}(i)}^{(s',l')} .
            \label{eq:compatibility}
        \end{equation}
        Eq.~\eqref{eq:compatibility} follows directly from Eq.~\eqref{eq:bondvec}, as every bond is the difference of the positions of the two atoms it connects, and therefore holds irrespective of the Wyckoff position or species of those atoms.

        At a given strain $\epsilon$ and temperature $T$, integrating the equilibrium distribution $\rho(\{\mathbf b\},\epsilon,T)$ over the constraint surface for all bonds but one, gives the distribution of that bond.
        Denoting the corresponding thermal average by $\langle \cdot \rangle$, we define the single bond density of shell $s$ of Wyckoff position $\kappa$ as
        \begin{equation}
            \rho_{\kappa,s}(\mathbf b,\epsilon,T) \equiv \left\langle \delta\!\left(\mathbf{G}_{\kappa,s,l}^{-1}\mathbf b_{\kappa}^{(s,l)} - \mathbf b\right)\right\rangle,
            \label{eq:singlebonddensity}
        \end{equation}
        which, by the symmetry equivalence established in Sec.~\ref{sec:theory-symmetry}, is independent of the choice of atom $i$ in $\kappa$, and neighbor bond $l$ in $s$.
        We note that $\rho_{\kappa,s}(\mathbf b,\epsilon,T)$ includes the correlations of every bond in the crystal, as they are coupled by the compatibility constraint defined in Eq.~\eqref{eq:compatibility}.

        The expected potential energy per atom of Wyckoff position $\kappa$ can then be written as
        \begin{equation}
            \mathcal{U}_\kappa(\epsilon,T) \equiv \frac{1}{2}\sum_{s=1}^{S_\kappa} m_{\kappa,s} \int V_{\kappa,s}(\mathbf b)\,\rho_{\kappa,s}(\mathbf b,\epsilon,T)\,d\mathbf b,
            \label{eq:U_kappa}
        \end{equation}
        with the expected potential energy per atom of the crystal given by the atomic-fraction-weighted sum over Wyckoff positions
        \begin{equation}
            \mathcal{U}(\epsilon,T) = \sum_{\kappa=1}^{K} c_\kappa\, \mathcal{U}_\kappa(\epsilon,T).
            \label{eq:U_total}
        \end{equation}
        The anharmonic contribution to the internal energy can be obtained by subtracting the equipartition potential energy of a classical harmonic crystal,
        \begin{equation}
            \mathcal{U}^{\mathrm{ah}}(\epsilon,T) \equiv \mathcal{U}(\epsilon,T) - \frac{3}{2}k_{\mathrm B}T,
            \label{eq:Uah_total}
        \end{equation}
        where each $V_{\kappa,s}$ in Eq.~\eqref{eq:U_kappa} is measured relative to its value at the strained equilibrium bond $\mathbf a^\epsilon_{\kappa,s}$, such that $\mathcal{U}^{\mathrm{ah}}(\epsilon,T)$ vanishes as $T \to 0$.
        The anharmonic free energy then follows directly as
        \begin{equation}
            \mathcal{F}^{\mathrm{ah}}(\epsilon,T) \equiv -T\int_0^T \frac{\mathcal{U}^{\mathrm{ah}}(\epsilon,T')}{(T')^2}\,dT' .
            \label{eq:Fah_total}
        \end{equation}
        For $K=1$, $S_1=1$, Eqs.~\eqref{eq:U_kappa}--\eqref{eq:Fah_total} recover Eqs.~(5) and (7) of Ref.~\cite{Swinburne2020} exactly.
        
        \subsection{The fully correlated mean-field approximation}
        \label{sec:theory-meanfield}
        
        Imposing the following mean-field approximation on the compatibility constraint prevents any two bonds of a shell originating from the same atom from moving independently:
        \begin{equation}
            \mathbf b_\kappa^{(s,l)} - \mathbf b_\kappa^{(s,l')} = \mathbf a^\epsilon_{\kappa,s,l} - \mathbf a^\epsilon_{\kappa,s,l'}.
            \label{eq:meanfield}
        \end{equation}
        This means that fixing the deviation of one bond from its equilibrium value fully determines every other bond in the shell.
        Comparisons against bond statistics from MD simulations showed that the above approximation was too restrictive, particularly for antiparallel bond pairs, motivating an analytical correction for their correlation~\cite{Swinburne2020}.
        Although this correction extends to any shell with an antiparallel bond pair, correlations between the reference bond of the shell and all other bonds in the crystal remain unaccounted for.
        We therefore replace Eq.~\eqref{eq:meanfield} with a covariance-derived coupling that quantifies the response of every bond in the crystal to a change in the reference bond of a shell.
        
        Let $\delta\mathbf b \equiv \mathbf b - \mathbf a^\epsilon$ denote the deviation of a bond from its strained equilibrium value.
        We approximate the deviation of the bond $\mathbf b_i^{(s',l')}$ in the crystal as a linear function of the deviation of the reference bond $\mathbf b_{\kappa,s}$, such that
        \begin{equation}
            \delta\mathbf b_i^{(s',l')} \approx \left\langle \delta\mathbf b_i^{(s',l')} \mid \delta\mathbf b_{\kappa,s} \right\rangle = \boldsymbol\alpha_{\kappa,s,i,s',l'}\,\delta\mathbf b_{\kappa,s},
            \label{eq:condmean}
        \end{equation}
        where $\boldsymbol\alpha_{\kappa,s,i,s',l'}$ is the corresponding normalized covariance matrix, which we derive in Sec.~\ref{sec:theory-covariance}.
        Eq.~\eqref{eq:condmean} gives the exact conditional expectation for a joint Gaussian distribution of harmonic bond fluctuations~\cite{Anderson2003, Bishop2006}, approximated here by discarding the associated variance.
        Unlike Eq.~\eqref{eq:meanfield}, this coupling is not restricted to bonds of a shared atom, but extends to any bond with a non-negligible covariance to the reference bond, including bonds in other shells of the same atom, bonds of other atoms of the same Wyckoff position, and bonds of atoms belonging to entirely different Wyckoff positions.
        
\begin{figure}[b]
            \includegraphics[width=1.0\columnwidth]{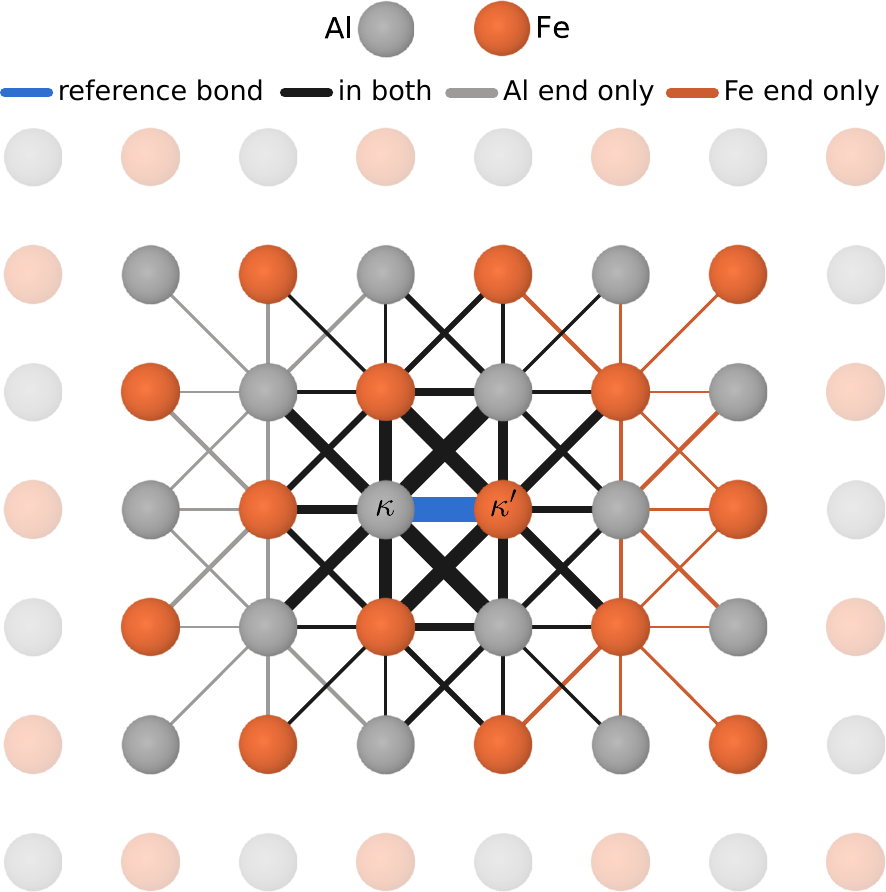}
            \caption{Bond-centered cluster of $\tilde V_{\kappa,s}$ (Eq.~\eqref{eq:Veff}) for the Al--Fe reference bond (blue) of a two-dimensional B2 analogue with two neighbor shells. Black bonds are reached from both ends, grey and orange bonds from the Al or Fe end only.
            Bond widths scale with $\|\boldsymbol\alpha\|$ (Eq.~\eqref{eq:alpha}). Faded atoms lie outside the cluster and are not accounted for in the sum.}
            \label{fig:cluster}
        \end{figure}

        Incorporating Eq.~\eqref{eq:condmean} and the symmetry operations $\mathbf G_{i,s',l'}$ of Sec.~\ref{sec:theory-symmetry} generalizes the shared motion defined by Eq.~\eqref{eq:meanfield} to arbitrary shells and atoms
        \begin{equation}
            \tilde{\mathbf b}_i^{(s',l')} \equiv \mathbf{G}_{i,s',l'}^{-1}\boldsymbol\alpha_{\kappa,s,i,s',l'}\big(\mathbf b - \mathbf a^\epsilon_{\kappa,s}\big) + \mathbf a^\epsilon_{\kappa(i),s'},
            \label{eq:predbond}
        \end{equation}
        which, when substituted into Eq.~\eqref{eq:energy}, gives the fully correlated mean-field effective potential for shell $s$ of Wyckoff position $\kappa$,
        \begin{equation}
            \tilde V_{\kappa,s}(\mathbf b,\epsilon) = \frac{1}{2}\sum_{\kappa'=1}^{K}\sum_{i=1}^{N_{\kappa'}} \sum_{s'=1}^{S_{\kappa'}} \sum_{l'=1}^{m_{\kappa',s'}} V_{\kappa',s'}\!\left(\tilde{\mathbf b}_i^{(s',l')}\right),
            \label{eq:Veff}
        \end{equation}
        summed formally over every bond in the crystal.
        In practice, we restrict this sum to the bonds of a cluster formed by the two atoms of the reference bond and their neighbors within the retained shells (Fig.~\ref{fig:cluster}).
        A bond with both ends in this cluster is enumerated from either end and enters with weight $1/2$, as in Eq.~\eqref{eq:energy}, whereas a bond extending beyond the cluster is enumerated once and assigned a weight of $1$, ensuring that every physical bond is counted exactly once.
        Since the clusters of the two shells of a reciprocal pair consist of the same atoms seen from opposite ends, their effective potentials satisfy $\tilde V_{\kappa',s'}(-\mathbf Q_{\kappa,s}\mathbf b,\epsilon) = \tilde V_{\kappa,s}(\mathbf b,\epsilon)$, such that a shared bond is described identically from either end.

        The mean-field single bond density follows as the Boltzmann inverse of $\tilde V_{\kappa,s}$,
        \begin{equation}
            \rho_{\kappa,s}(\mathbf b,\epsilon,T) \equiv \frac{1}{Z_{\kappa,s}(\epsilon,T)}\exp\!\left[-\frac{\tilde V_{\kappa,s}(\mathbf b,\epsilon)}{k_{\mathrm B}T}\right],
            \label{eq:mfdensity}
        \end{equation}
        with $Z_{\kappa,s}(\epsilon,T) \equiv \int \exp\!\left[-\tilde V_{\kappa,s}(\mathbf b',\epsilon)/k_{\mathrm B}T\right]d\mathbf b'$, generalizing Eq.~(15) of Ref.~\cite{Swinburne2020}.
        
        Retaining in Eq.~\eqref{eq:Veff} only the bonds of shell $s$ enumerated from the representative atom, $i=i_\kappa$ and $s'=s$, with $\boldsymbol\alpha_{\kappa,s,i_\kappa,s,l'} = \mathbb I$ for every $l' \in \{1, \dots, m_{\kappa,s}\}$, reduces the expression identically to the mean-field effective potential in Eq.~(12) of Ref.~\cite{Swinburne2020}.
        The uncorrelated mean-field bond model can therefore be recovered as the trivial limit of Eq.~\eqref{eq:condmean}, restricted to a single shell of a single atom.
        The correlated refinement in Eqs.~(13)--(15) of Ref.~\cite{Swinburne2020}, however, is not a special case of the same form, as it is derived from an exact compatibility solution for the antiparallel bond pair, rather than from a coupling coefficient.
        
        \subsection{Bond pair covariances from the Hessian}
        \label{sec:theory-covariance}

        We determine the normalized covariance matrices $\boldsymbol\alpha_{\kappa,s,i,s',l'}$ introduced in Sec.~\ref{sec:theory-meanfield} from the Hessian of the total energy $U$ defined in Eq.~\eqref{eq:energy}, expressed here as a function of atomic positions as
        \begin{equation}
            \boldsymbol\Phi_{ij} \equiv \left.\frac{\partial^2 U}{\partial \mathbf R_i\,\partial \mathbf R_j}\right|_{\{\mathbf R\}=\{\mathbf R^\epsilon\}},
            \label{eq:hessian}
        \end{equation}
        where $\boldsymbol\Phi$ is the mass-unweighted force-constant matrix. 
        For simplicity, we consider the force constants of the relaxed crystal at zero temperature with strain $\epsilon=0$, and omit the isotropic strain label throughout this subsection.
        
        In the classical harmonic approximation, the canonical distribution of the atomic displacements about their equilibrium positions $\delta\mathbf R_i \equiv \mathbf R_i - \mathbf R_i^{0}$ is Gaussian, with covariance
        \begin{equation}
            \langle \delta\mathbf R_i \otimes \delta\mathbf R_j \rangle = k_{\mathrm B}T\,[\boldsymbol\Phi^{+}]_{ij},
            \label{eq:greensfunction}
        \end{equation}
        where the superscript $^{+}$ denotes the Moore-Penrose pseudo-inverse~\cite{Penrose1955}, as $\boldsymbol\Phi$ is singular under a uniform translation of the crystal.
        Since the direct inversion of the $3N\times 3N$ matrix $\boldsymbol\Phi$ is impractical, we exploit the periodicity of the relaxed crystal, whose primitive cell contains $B \ge K$ basis sites $\mu(i) \in \{1, \dots, B\}$.
        Using Bloch's theorem to block-diagonalize Eq.~\eqref{eq:hessian} into a $3B\times 3B$ dynamical matrix $\mathbf D(\mathbf q)$ per wavevector $\mathbf q$, we obtain, for a fixed atom $i$ at basis site $\mu$,
        \begin{equation}
            \mathbf D_{\mu\mu'}(\mathbf q) = \sum_{j\,:\,\mu(j)=\mu'} \boldsymbol\Phi_{ij}\, e^{i\mathbf q\cdot(\mathbf R_j - \mathbf R_i)}.
            \label{eq:dynmat}
        \end{equation}
        This is independent of the choice of $i$, since any atom with $\mu(i)=\mu$ gives the same result by translational symmetry.
        Unlike the conventional phonon dynamical matrix, Eq.~\eqref{eq:dynmat} is not mass-weighted, allowing Eq.~\eqref{eq:greensfunction} to be rewritten as
        \begin{equation}
            \langle \delta\mathbf R_i \otimes \delta\mathbf R_j \rangle = \frac{k_{\mathrm B}T}{N_c}\sum_{\mathbf q} \mathbf D^{+}_{\mu(i)\mu(j)}(\mathbf q)\,e^{i\mathbf q\cdot(\mathbf R_i-\mathbf R_j)},
            \label{eq:posvariance}
        \end{equation}
        where $N_c$ is the number of primitive cells.

        Since $\boldsymbol\Phi$ is evaluated on a finite periodic supercell, the sum over $\mathbf q$ in Eq.~\eqref{eq:posvariance} runs over the wavevectors commensurate with that supercell and is exact, requiring no separately converged wavevector mesh~\cite{Togo2023}.
        The dynamical matrix $\mathbf D(\mathbf q)$ is singular only at $\mathbf q = 0$, where its null space corresponds to uniform translations of the crystal, and is removed by the pseudo-inverse~\cite{Penrose1955}.
        As bond deviations are differences of displacements, they remain unaffected by this removal.
        Eq.~\eqref{eq:posvariance} is also independent of the atomic masses for any pair of basis sites $(\mu(i),\mu(j))$, and therefore applies to crystals with any number of chemical species without modification (Appendix~\ref{app:mass-cancellation}).

        We denote by $r_1 \equiv i_\kappa$ and $r_2 \equiv n_{s,1}(i_\kappa)$ the two atoms forming the reference bond $\mathbf b_{\kappa,s}$, and by $i_1 \equiv i$ and $i_2 \equiv n_{s',l'}(i)$ the two atoms forming the bond $\mathbf b_{i,s',l'}$, where $l'$ ranges over every neighbor in shell $s'$.
        The bond deviations are then the differences of the atomic displacements
        \begin{equation}
            \delta\mathbf b_{\kappa,s} = \delta\mathbf R_{r_2} - \delta\mathbf R_{r_1}; \quad
            \delta\mathbf b_{i,s',l'} = \delta\mathbf R_{i_2} - \delta\mathbf R_{i_1},
        \label{eq:bonddiff}
        \end{equation}
        and their covariance follows by bilinearity as
        \begin{widetext}
            \begin{equation}
                \mathrm{Cov}\big(\delta\mathbf b_{i,s',l'}, \delta\mathbf b_{\kappa,s}\big) =
                \big\langle \delta\mathbf R_{i_2} \otimes \delta\mathbf R_{r_2} \big\rangle
                - \big\langle \delta\mathbf R_{i_2} \otimes \delta\mathbf R_{r_1} \big\rangle
                - \big\langle \delta\mathbf R_{i_1} \otimes \delta\mathbf R_{r_2} \big\rangle
                + \big\langle \delta\mathbf R_{i_1} \otimes \delta\mathbf R_{r_1} \big\rangle,
                \label{eq:bondcov}
            \end{equation}
        \end{widetext}
        with each term evaluated from Eq.~\eqref{eq:posvariance}.
        Normalizing by the self-covariance of the reference bond $\mathbf C_{\kappa,s} \equiv \mathrm{Cov}(\delta\mathbf b_{\kappa,s}, \delta\mathbf b_{\kappa,s})$ gives
        \begin{equation}
            \boldsymbol\alpha_{\kappa,s,i,s',l'} = \mathrm{Cov}\big(\delta\mathbf b_{i,s',l'}, \delta\mathbf b_{\kappa,s}\big)\,\mathbf C_{\kappa,s}^{-1},
            \label{eq:alpha}
        \end{equation}
        which turns the covariance into a coefficient matrix that predicts a linear response of any bond to a unit deviation of the reference bond.
        Every term in Eq.~\eqref{eq:bondcov} carries the same factor of $k_{\mathrm B}T$, which cancels identically in Eq.~\eqref{eq:alpha}.
        The matrix $\boldsymbol\alpha_{\kappa,s,i,s',l'}$ can thus be computed exactly from the force constants of a given supercell, independent of temperature and atomic mass.
        Since every bond deviation is a difference of atomic displacements, the deviations predicted by Eq.~\eqref{eq:condmean} are differences of a single displacement field corresponding to the conditional mean of the atomic displacements at fixed $\delta\mathbf b_{\kappa,s}$, and satisfy Eq.~\eqref{eq:compatibility} identically.
        For harmonic bonding potentials and an untruncated sum over bonds, the same displacement field is the minimum-energy configuration of the crystal at fixed $\delta\mathbf b_{\kappa,s}$, such that $\tilde V_{\kappa,s}$ reduces to $\tfrac{1}{2}k_{\mathrm B}T\,\delta\mathbf b^{\top}\mathbf C_{\kappa,s}^{-1}\delta\mathbf b$ and the mean-field single bond density recovers the exact harmonic single bond density.
        The fully correlated mean-field approximation is therefore exact for harmonic bonding potentials.
        For anharmonic ones, each bonding potential still enters $\tilde V_{\kappa,s}$ in full, and the approximation introduces error only through the harmonic coupling between bonds, the truncation of the sum over bonds in Eq.~\eqref{eq:Veff}, and away from $\epsilon=0$, the strain dependence of the covariances.
        
        \subsection{Thermodynamic self-consistency}
        \label{sec:theory-selfconsistency}
        
        The mean-field single bond density $\rho_{\kappa,s}(\mathbf b,\epsilon,T)$ is not guaranteed to satisfy the volume constraint by construction.
        Since an anharmonic bonding potential is asymmetric about its minimum, the mean bond $\langle\mathbf b\rangle_{\kappa,s}$ predicted by the fully correlated mean-field effective potential $\tilde V_{\kappa,s}(\mathbf b,\epsilon)$ is, in general, displaced from the equilibrium bond $\mathbf a^\epsilon_{\kappa,s}$ imposed by the strain.
        Ref.~\cite{Swinburne2020} enforced the volume constraint through a single Lagrange multiplier, which we generalize here by adding a multiplier $\lambda_{\kappa,s}(\epsilon,T)$ to the effective potential of every shell $s$ of every Wyckoff position $\kappa$, such that Eq.~\eqref{eq:Veff} becomes
        \begin{equation}
            \tilde V_{\kappa,s}(\mathbf b,\epsilon) \;\to\; \tilde V_{\kappa,s}(\mathbf b,\epsilon) + \lambda_{\kappa,s}(\epsilon,T)\,\hat{\mathbf a}_{\kappa,s}\cdot\mathbf b,
            \label{eq:lagrange}
        \end{equation}
        with $\hat{\mathbf a}_{\kappa,s} \equiv \mathbf a^\epsilon_{\kappa,s}/|\mathbf a^\epsilon_{\kappa,s}|$.
        Each $\lambda_{\kappa,s}(\epsilon,T)$ is then solved to satisfy
        \begin{equation}
            \langle \mathbf b\rangle_{\kappa,s} \cdot \hat{\mathbf a}_{\kappa,s} = |\mathbf a^\epsilon_{\kappa,s}|,
            \label{eq:volumeconstraint}
        \end{equation}
        enforcing the volume constraint across all shells.
        
        Eq.~\eqref{eq:volumeconstraint}, however, says nothing about whether $\rho_{\kappa,s}(\mathbf b,\epsilon,T)$ reproduces the target temperature $T$.
        The contribution of shell $s$ to the virial temperature of Wyckoff position $\kappa$ is given by
        \begin{equation}
            T^{\mathrm{vir}}_{\kappa,s}(\epsilon,T') \equiv \frac{m_{\kappa,s}}{2dk_{\mathrm B}}\,\big\langle \delta\mathbf b\cdot\nabla V_{\kappa,s}(\mathbf b)\big\rangle_{\kappa,s},
            \label{eq:Tvir_shell}
        \end{equation}
        where $\langle\cdot\rangle_{\kappa,s}$ denotes the average over $\rho_{\kappa,s}(\mathbf b,\epsilon,T')$ and $\delta\mathbf b = \mathbf b - \mathbf a^\epsilon_{\kappa,s}$, with the virial temperature of the representative atom of $\kappa$ obtained by summing over its shells as
        \begin{equation}
            T^{\mathrm{vir}}_{\kappa}(\epsilon,T') \equiv \sum_{s=1}^{S_\kappa} T^{\mathrm{vir}}_{\kappa,s}(\epsilon,T').
            \label{eq:Tvir_kappa}
        \end{equation}
        We solve for a single effective temperature $\tilde T$ at which every $\rho_{\kappa,s}(\mathbf b,\epsilon,\tilde T)$ is evaluated, such that the atomic-fraction-weighted virial temperature of the crystal recovers the target temperature as
        \begin{equation}
            \sum_{\kappa=1}^{K} c_\kappa\, T^{\mathrm{vir}}_{\kappa}(\epsilon,\tilde T) = T,
            \label{eq:Teff}
        \end{equation}
        which, for $K=1$, reduces to the virial-corrected mean-field bond model of Ref.~\cite{Swinburne2020}.
        We note that Eq.~\eqref{eq:Teff} constrains only the crystal average, with the virial temperatures of individual Wyckoff positions running above or below $T$.
        The equipartition term subtracted in Eq.~\eqref{eq:Uah_total} is therefore evaluated per Wyckoff position as $3/2k_{\mathrm B}T^{\mathrm{vir}}_{\kappa}(\epsilon,\tilde T)$, whose atomic-fraction-weighted sum recovers $3/2k_{\mathrm B}T$ from Eq.~\eqref{eq:Teff}.
        
        The pressure contribution of a shell $s$ to Wyckoff position $\kappa$ follows, as
        \begin{equation}
            P_{\kappa,s}(\epsilon',T') \equiv -\frac{m_{\kappa,s}P_0(\epsilon')}{2}\,\big\langle \mathbf a^{\epsilon'}_{\kappa,s}\cdot\nabla V_{\kappa,s}(\mathbf b)\big\rangle_{\kappa,s},
            \label{eq:Pshell}
        \end{equation}
        where $P_0(\epsilon') \equiv N/[dV_0(1+\epsilon')^d]$, which, when summed over shells and atomic-fraction-weighted over Wyckoff positions gives the pressure of the crystal
        \begin{equation}
            P(\epsilon,T) = \sum_{\kappa=1}^{K} c_\kappa \sum_{s=1}^{S_\kappa} P_{\kappa,s}(\epsilon,\tilde T).
            \label{eq:Ptotal}
        \end{equation}
        Together, Eqs.~\eqref{eq:volumeconstraint} and~\eqref{eq:Teff} generalize the self-consistency condition of the isochoric (NVT) ensemble, given by Eq.~(18) of Ref.~\cite{Dsouza2024}, to multiple Wyckoff positions and shells.
        In the isobaric (NPT) ensemble, the strain $\epsilon$ is additionally solved such that $P(\epsilon,T)$ matches the target pressure, with Eqs.~\eqref{eq:volumeconstraint} and~\eqref{eq:Teff} re-solved at every trial value of $\epsilon$, generalizing the NPT condition given by Eq.~(20) of Ref.~\cite{Dsouza2024}.
        
        Combined with the bond-centered effective potentials of Sec.~\ref{sec:theory-meanfield}, a single effective temperature ensures that the two shells of a reciprocal pair describe their shared bonds identically.
        Since $\tilde V_{\kappa',s'}(-\mathbf Q_{\kappa,s}\mathbf b,\epsilon) = \tilde V_{\kappa,s}(\mathbf b,\epsilon)$ and $\hat{\mathbf a}_{\kappa',s'} = -\mathbf Q_{\kappa,s}\hat{\mathbf a}_{\kappa,s}$, Eq.~\eqref{eq:volumeconstraint} gives $\lambda_{\kappa',s'} = \lambda_{\kappa,s}$, and the two densities satisfy $\rho_{\kappa',s'}(-\mathbf Q_{\kappa,s}\mathbf b,\epsilon,\tilde T) = \rho_{\kappa,s}(\mathbf b,\epsilon,\tilde T)$.
        Following Eq.~\eqref{eq:reciprocity} and the bond counts of Sec.~\ref{sec:theory-symmetry}, the shared bonds also contribute equally to the virial temperature and pressure from either end.
        Crystals with a single Wyckoff position obey the same conditions without requiring any special treatment.
    
    \section{Methodology}
    \label{sec:methodology}
    
        \subsection{Choice of crystal structures}
        \label{sec:methods-structures}
    
        To demonstrate each feature of the generalization, we select six crystal structures of increasing complexity (Table~\ref{tab:structures}), each isolating one or more features.
        Among the unary systems, fcc Al (A1) connects the generalization to the original formulation, which has been previously validated against MD for fcc crystals~\cite{Swinburne2020, Dsouza2024}.
        Its anharmonicity is dominated by the first nearest-neighbor shell, and its further shells carry no bonding potential of their own, testing the coupling of Sec.~\ref{sec:theory-meanfield} alone.
        bcc Fe (A2) also has a single Wyckoff position and a single chemical species, but isolates the contribution of multiple shells carrying their own bonding potentials, in a lattice where next-nearest neighbors have non-negligible contributions~\cite{Dsouza2025thesis}.
        hcp Mg (A3) adds a second basis site within a single Wyckoff position, testing the reduction of every atom to its representative through the symmetry operations of Sec.~\ref{sec:theory-symmetry}.
        The binary system AlFe (B2) has two Wyckoff positions with different chemical species that share a single shell of bonds, the minimal case of a reciprocal pair joining two Wyckoff positions described in Sec.~\ref{sec:theory-selfconsistency}.
        AlFe$_3$ (D0$_3$) has three Wyckoff positions but only two chemical species, requiring the model to keep the two Fe sites distinct from each other, while its shared bonds couple all three positions.
        The ternary system Al$_2$FeNi (L2$_1$) is a Heusler alloy~\cite{Graf2011} that combines three chemical species with three Wyckoff positions, putting every feature of the generalization to test at once.
        The six crystal structures are parametrized from five different reference potentials, four of the embedded-atom method (EAM)~\cite{Daw1984} and one angular-dependent potential (ADP)~\cite{Mishin2005}, providing bonding potentials of different stiffness and anharmonicity.
        \begin{table*}
            \caption{Crystal structures used to validate the generalized mean-field bond model. Each system is chosen to isolate a distinct feature of the generalization.}
            \label{tab:structures}
            \begin{ruledtabular}
            \begin{tabular}{lllll}
            Formula & Strukturbericht & Space group & Wyckoff positions & Reference potential \\
            \hline
            Al & A1 (fcc) & $Fm\bar3m$ & Al: $4a$ & Mendelev \textit{et al.}~\cite{Mendelev2008} (EAM) \\
            Fe & A2 (bcc) & $Im\bar3m$ & Fe: $2a$ & Starikov \textit{et al.}~\cite{Starikov2021} (ADP) \\
            Mg & A3 (hcp) & $P6_3/mmc$ & Mg: $2c$ & Sun \textit{et al.}~\cite{Sun2006} (EAM) \\
            AlFe & B2 & $Pm\bar3m$ & Al: $1a$, Fe: $1b$ & Mendelev \textit{et al.}~\cite{Mendelev2005} (EAM) \\
            AlFe$_3$ & D0$_3$ & $Fm\bar3m$ & Al: $4a$, Fe: $4b$, Fe: $8c$ & Mendelev \textit{et al.}~\cite{Mendelev2005} (EAM) \\
            Al$_2$FeNi & L2$_1$ (Heusler) & $Fm\bar3m$ & Al: $8c$, Ni: $4b$, Fe: $4a$ & Hayakawa and Xu~\cite{Hayakawa2024} (EAM) \\
            \end{tabular}
            \end{ruledtabular}
        \end{table*}

        \subsection{Parametrization of bonding potentials}
        \label{sec:methods-parametrization}
    
        While the mean-field bond model is agnostic to the functional form of the bonding potentials, we favor in this work one that is fast, simple, and easy to reproduce, rather than the flexible and more accurate bonding potentials fit directly to MLIPs in Ref.~\cite{Dsouza2024} using the local anharmonic approximation~\cite{Glensk2015}.
        To this end, we choose central Morse-type potentials of the form
        \begin{equation}
            V(r) = D_e\left(1-e^{-\alpha(r-r_0)}\right)^2,
            \label{eq:morse}
        \end{equation}
        with $r = |\mathbf b|$, which is anharmonic by construction and defined by only three parameters per shell.
        The parameters of each shell are calibrated from the force constants of the corresponding empirical interatomic potential in Table~\ref{tab:structures}, computed by finite displacements of $0.01\,\text{\AA}$ in the simulation cells given in Table~\ref{tab:computational}.
        The curvature of each shell is taken as the longitudinal projection of the force-constant block of its bonds, $k = -\hat{\mathbf a}_{\kappa,s,l}\cdot\boldsymbol\Phi_{i\,n_{s,l}(i)}\cdot\hat{\mathbf a}_{\kappa,s,l}$, averaged over the bonds of the shell, while the transverse components of $\boldsymbol\Phi$ are discarded.
        The equilibrium bond length $r_0$ is kept fixed to the reference bond length of each shell, leaving $D_e$ and $\alpha$ as the two free parameters.
        A single force-constant matrix, however, fixes only the curvature $k \equiv V''(r_0) = 2D_e\alpha^2$ at that one volume, resulting in one equation for two unknowns.
    
        We therefore evaluate additional Hessians at isotropic strains $\epsilon \in [-0.07, 0.07]$, sampling the curvature $k$ at different bond lengths.
        The curvature around $r_0$ then follows
        \begin{equation}
            k(r) = 2D_e\alpha^2\left(2e^{-2\alpha(r-r_0)}-e^{-\alpha(r-r_0)}\right),
            \label{eq:curvature}
        \end{equation}
        giving the closed-form estimates $\alpha_0 = -(dk/dr)/(3k)$ and $D_{e,0} = k/(2\alpha_0^2)$, with $k$ measured directly from the reference-volume force constants, and $dk/dr$ from a linear least-squares fit of $k$ against bond lengths across the strained volumes.
        We use this estimate to initialize a nonlinear least-squares fit of Eq.~\eqref{eq:curvature} to the sampled $(r,k)$ pairs, giving the parameters $D_e$ and $\alpha$ for each shell.
        Since both ends of a reciprocal pair share the same force-constant block and bond length, the fitted parameters satisfy the reciprocity of Eq.~\eqref{eq:reciprocity}.
        A shell whose $k$ is negative at the reference volume corresponds to an anti-restoring direction rather than a genuine potential well, and is assigned no potential ($D_e=\alpha=0$) rather than fit to an unphysical result.
        The fitted parameters for every shell and Wyckoff position for each of the six structures are provided in the accompanying dataset~\cite{Dsouza2026data}.
    
        \subsection{Computational details}
        \label{sec:methods-computational}
    
        The symmetry datasets used in this work were obtained from the spglib~\cite{Togo2024} package, and all MD simulations were performed using LAMMPS~\cite{Thompson2022}.
        The generalized mean-field bond model has been implemented in the accompanying code~\cite{Dsouza2026code}.
        The cluster defined in Sec.~\ref{sec:theory-meanfield} extends the neighbors of each end of the reference bond by as many shells as are retained, and bonds whose covariances $\|\boldsymbol\alpha_{\kappa,s,i,s',l'}\| < 10^{-3}$ are discarded.
        Shells that are related by symmetry or by reciprocity and share a single bond density are solved only once.
        The effective potential is evaluated on a grid of $15\times15\times15$ points and interpolated by cubic splines onto $51\times51\times51$ points, on which the single bond densities are evaluated.
        
        The neighbor shells, simulation cells, and total atom counts for each structure are given in Table~\ref{tab:computational}.
        We use the same simulation cell to compute the force constants and to perform the MD simulations to ensure that the model and MD describe the same finite periodic crystal, with the same commensurate wavevectors entering Eq.~\eqref{eq:posvariance}.
        For every structure, the retained shells as well as the further shells of the bond-centered clusters lie well within half the length of the simulation cell, such that no bond or force-constant block couples an atom to its own periodic image, and the parametrization of the bonding potentials is unaffected by the size of the cell.
        Since the covariances and the MD simulations share the same supercell, the comparison between them is free of finite-size effects.
        \begin{table}
            \caption{Neighbor shells, counted as site-symmetry orbits, simulation cell used for both the Hessians and the MD simulations, and total atom count for each structure.}
            \label{tab:computational}
            \begin{ruledtabular}
            \begin{tabular}{lccc}
            Formula & Neighbor shells & Simulation cell & Atoms \\
            \hline
            Al & 3 & $5\times5\times5$ & 500 \\
            Fe & 3 & $6\times6\times6$ & 432 \\
            Mg & 3 & $5\times5\times5$ & 500 \\
            AlFe & 2 & $6\times6\times6$ & 432 \\
            AlFe$_3$ & 2 & $4\times4\times4$ & 1024 \\
            Al$_2$FeNi & 2 & $4\times4\times4$ & 1024 \\
            \end{tabular}
            \end{ruledtabular}
        \end{table}

        For Mg, the $c/a$ ratio was fixed at its zero-temperature value, as the isotropic strain of Sec.~\ref{sec:theory-symmetry} does not independently constrain the two lattice parameters of a hexagonal cell.
        Because the EAM-predicted $c/a$ ratio departs from the ideal value (1.628 vs.\ the ideal 1.633), the twelve neighbors that would otherwise form a single degenerate shell in an ideal hcp lattice instead resolve into two distinct, closely spaced distances (3.178 and 3.184\,\text{\AA}), and are retained as separate shells.
    
        We used the same Morse-type bonding potentials for the MD simulations as in the generalized mean-field bond model rather than the reference potentials in Table~\ref{tab:structures}, so that any disagreement reflected the accuracy of the model itself and not the truncation of the many-body reference potential.
        Each bond was assigned the bonding potential of its shell through a fixed bond topology, to ensure that the MD simulations sampled exactly the Hamiltonian of Eq.~\eqref{eq:energy}.
        Both the model and the MD reference were evaluated at ten temperatures evenly spaced between $100\,\mathrm{K}$ and $1000\,\mathrm{K}$.
        Each temperature was sampled with five independent runs of $10\,000$ equilibration and $25\,000$ production steps at a $1\,\mathrm{fs}$ timestep, using a Langevin thermostat for the NVT ensemble, and an additional isotropic Nos\'e-Hoover barostat for the NPT ensemble, to maintain zero pressure.

        The anharmonic internal energy per atom from the MD simulations was obtained by subtracting the harmonic contribution of the $3N-3$ vibrational degrees of freedom from the potential energy $\mathcal U^{\mathrm{ah}}_{\mathrm{MD}} = [U_{\mathrm{MD}} - 1/2(3N-3)k_{\mathrm B}T^{\mathrm{kin}}_{\mathrm{MD}}]/N$, where $T^{\mathrm{kin}}_{\mathrm{MD}}$ is the instantaneous kinetic temperature, and $-3$ corresponds to the center-of-mass correction.
        In the NPT ensemble, the static energy of the lattice at the sampled strain is additionally subtracted to keep the result consistent with Eq.~\eqref{eq:Uah_total}.
        Means and standard errors are obtained by taking the block average over each run to account for time correlations, and are combined over the five independent runs.
        Since Eq.~\eqref{eq:Fah_total} requires $\mathcal U^{\mathrm{ah}}_{\mathrm{MD}}$ to have values down to $T=0$, below the lowest sampled temperature, we fit the MD values by weighted least squares to $\mathcal U^{\mathrm{ah}}_{\mathrm{MD}}(T) = \sum_{n=2}^{4} a_n T^n$ with no constant term, as we expect $\mathcal U^{\mathrm{ah}}_{\mathrm{MD}}\to 0$ as $T\to0$, and exclude the linear term, as the harmonic contribution has already been removed.
        Eq.~\eqref{eq:Fah_total} then integrates analytically to $\mathcal F^{\mathrm{ah}}_{\mathrm{MD}}(T) = -\sum_{n=2}^{4} a_n T^n/(n-1)$, with its standard error propagated from the covariance of the fitted coefficients.
        $\mathcal F^{\mathrm{ah}}$ for the model is obtained by trapezoidal integration using Eq.~\eqref{eq:Fah_total} over the same temperatures, extrapolated linearly from the two lowest temperatures to $T=0$.

    \section{Results}
    \label{sec:results}
    
    Each of Figs.~\ref{fig:Al-thermo}--\ref{fig:Al2FeNi-thermo} summarizes one crystal structure, with its unit cell shown in panel (a).
    Panels (b)--(d) show the bond densities of the most relevant shells at $1000\,\mathrm{K}$ in the NVT ensemble, the highest temperature considered, and the one at which the anharmonic distortion of the densities is largest, with filled contours from MD and line contours from the converged generalized mean-field bond (MFB) model.
    Panels (e)--(h) show the anharmonic internal energy, the anharmonic free energy, the pressure, and the lattice expansion over the full temperature range respectively, with shaded regions marking the standard error of the MD reference over the five independent runs.
    The pressure is shown only for the NVT ensemble, in which the strain is fixed and the pressure is a prediction, and the lattice expansion only for the NPT ensemble, in which the strain is solved to match the target pressure and the expansion is a prediction.
    
    For all six structures, the MFB model reproduces the MD bond densities, and agrees with the MD thermodynamics over the full temperature range to within the standard error, with the largest deviations at the highest temperatures, but well below the $1\,\mathrm{meV/atom}$ threshold.
    The following subsections discuss each structure only where it tests a feature of the generalization beyond this agreement.
    
    In each of the structures, the anharmonic internal energy $\mathcal U^{\mathrm{ah}}$ in the NPT ensemble exceeds that in the NVT ensemble.
    Since every bonding potential is measured relative to its value at the strained equilibrium bond (Sec.~\ref{sec:theory-bondlattice}), $\mathcal U^{\mathrm{ah}}$ excludes the static energy of the strained lattice, and contains only the energy of thermal fluctuations in excess of the harmonic value.
    A harmonic crystal therefore gives $\mathcal U^{\mathrm{ah}} = 0$ at any volume, and both ensembles measure the explicit anharmonicity of the bonding potentials, at the fixed volume of the relaxed crystal in the NVT ensemble, and at the thermally expanded volume in the NPT ensemble.
    At $\epsilon = 0$ in the NVT ensemble, the bonds fluctuate about the minima of their respective bonding potentials where the anharmonicity is weakest, and $\mathcal U^{\mathrm{ah}}$ and $\mathcal F^{\mathrm{ah}}$ remain within $\pm 2\,\mathrm{meV/atom}$ for every structure.
    In the NPT ensemble, thermal expansion stretches the bonds beyond $r_0$, onto the softer and more asymmetric attractive regime of the bonding potential, where the same thermal fluctuations sample a considerably more anharmonic potential energy surface.
    
    We reiterate that the values presented in Figs.~\ref{fig:Al-thermo}--\ref{fig:Al2FeNi-thermo} reflect the anharmonicity of the Morse-type bonding potentials, parametrized only from the zero-temperature curvature of the reference interatomic potentials at several volumes (Sec.~\ref{sec:methods-parametrization}), and not that of the reference interatomic potentials themselves.
    The relevant comparison throughout is therefore the MFB model against MD with identical bonding potentials, not against any other simulation or experiment.
    The bond densities of every shell and Wyckoff position for each of the six crystal structures are provided in the accompanying dataset~\cite{Dsouza2026data}.
    
        \subsection{Unary systems}
        \label{sec:results:unary}
    
\begin{figure}[b]
            \includegraphics[width=1.0\columnwidth]{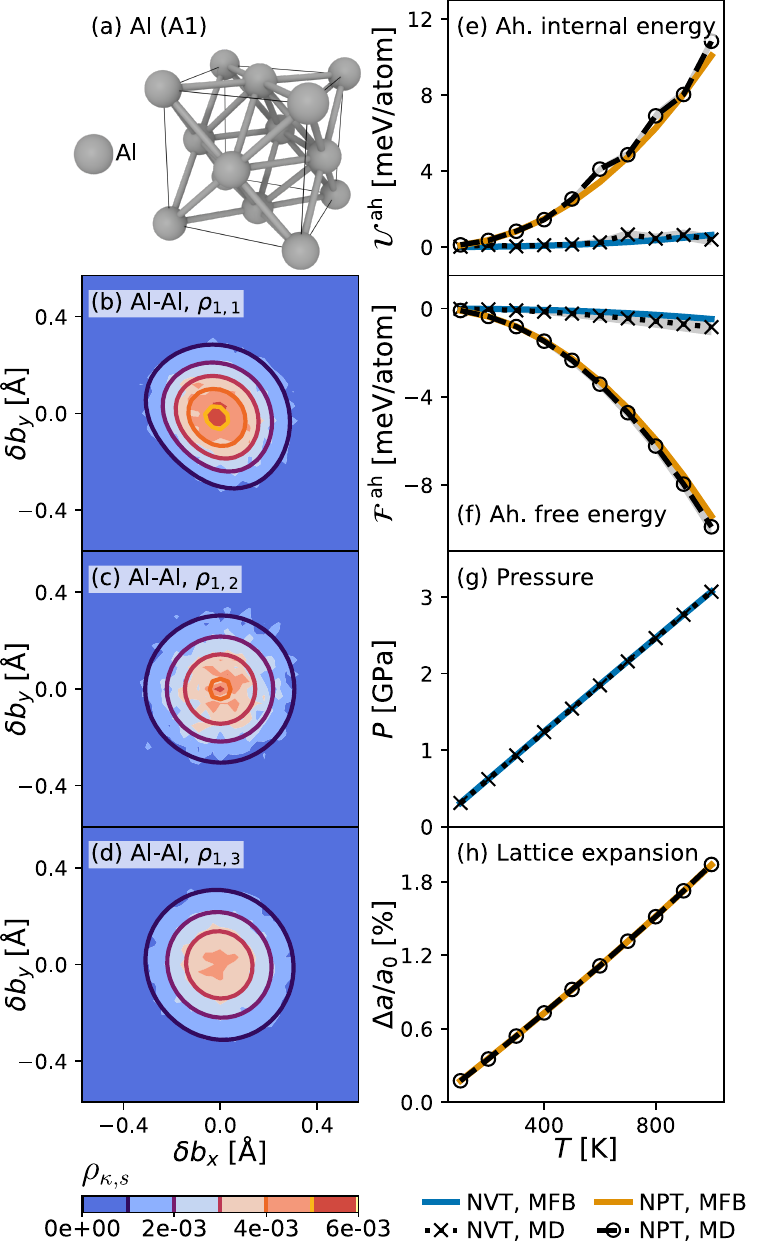}
            \caption{Bond densities and thermodynamics of Al. (a) Unit cell. (b)--(d) Al-Al bond density $\rho_{1,s}$ for shells $s=1,2,3$, filled contours from MD and line contours from the converged generalized mean-field bond (MFB) model, both at $1000\,\mathrm{K}$ in the NVT ensemble. (e) Anharmonic internal energy, (f) anharmonic free energy, (g) pressure, and (h) lattice expansion versus temperature, comparing the generalized model against MD in the NVT and NPT ensembles.}
            \label{fig:Al-thermo}
        \end{figure}

\begin{figure}[b]
            \includegraphics[width=1.0\columnwidth]{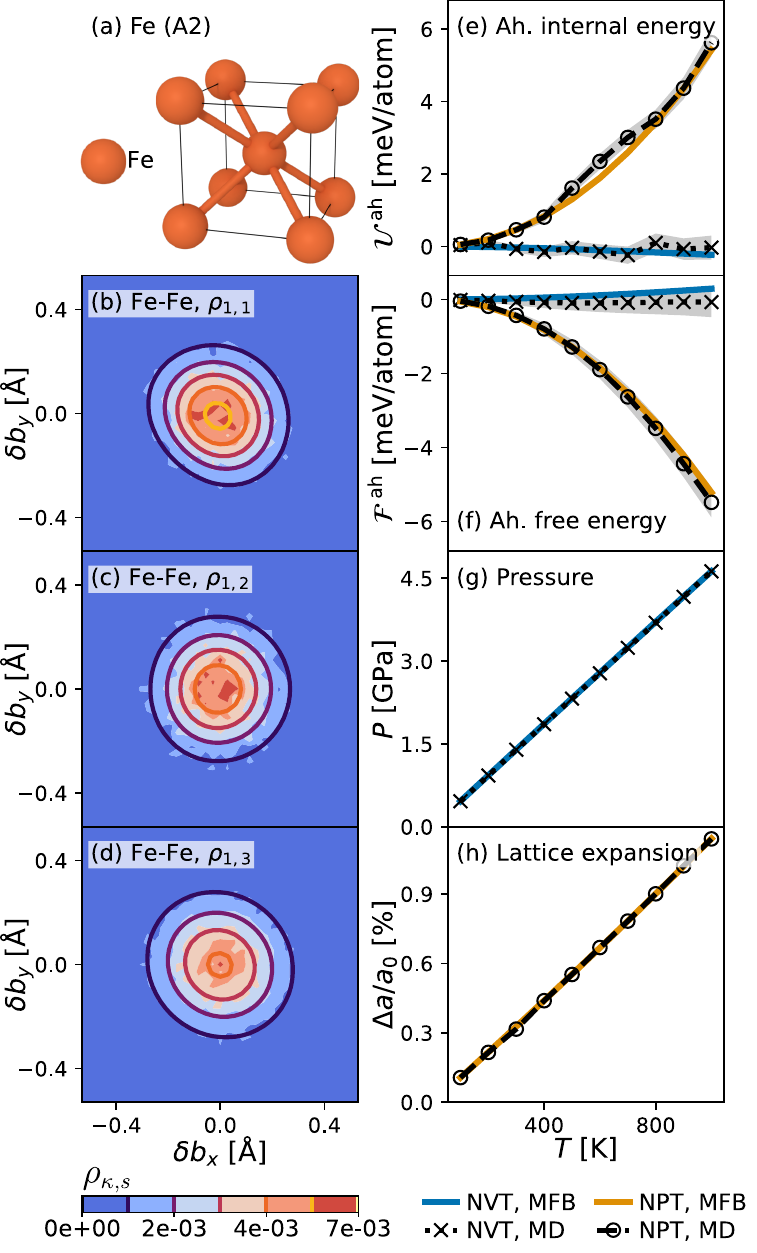}
            \caption{Bond densities and thermodynamics of Fe. (a) Unit cell. (b)--(d) Fe-Fe bond density $\rho_{1,s}$ for shells $s=1,2,3$, filled contours from MD and line contours from the converged generalized mean-field bond (MFB) model, both at $1000\,\mathrm{K}$ in the NVT ensemble. (e) Anharmonic internal energy, (f) anharmonic free energy, (g) pressure, and (h) lattice expansion versus temperature, comparing the generalized model against MD in the NVT and NPT ensembles.}
            \label{fig:Fe-thermo}
        \end{figure}

\begin{figure}[b]
            \includegraphics[width=1.0\columnwidth]{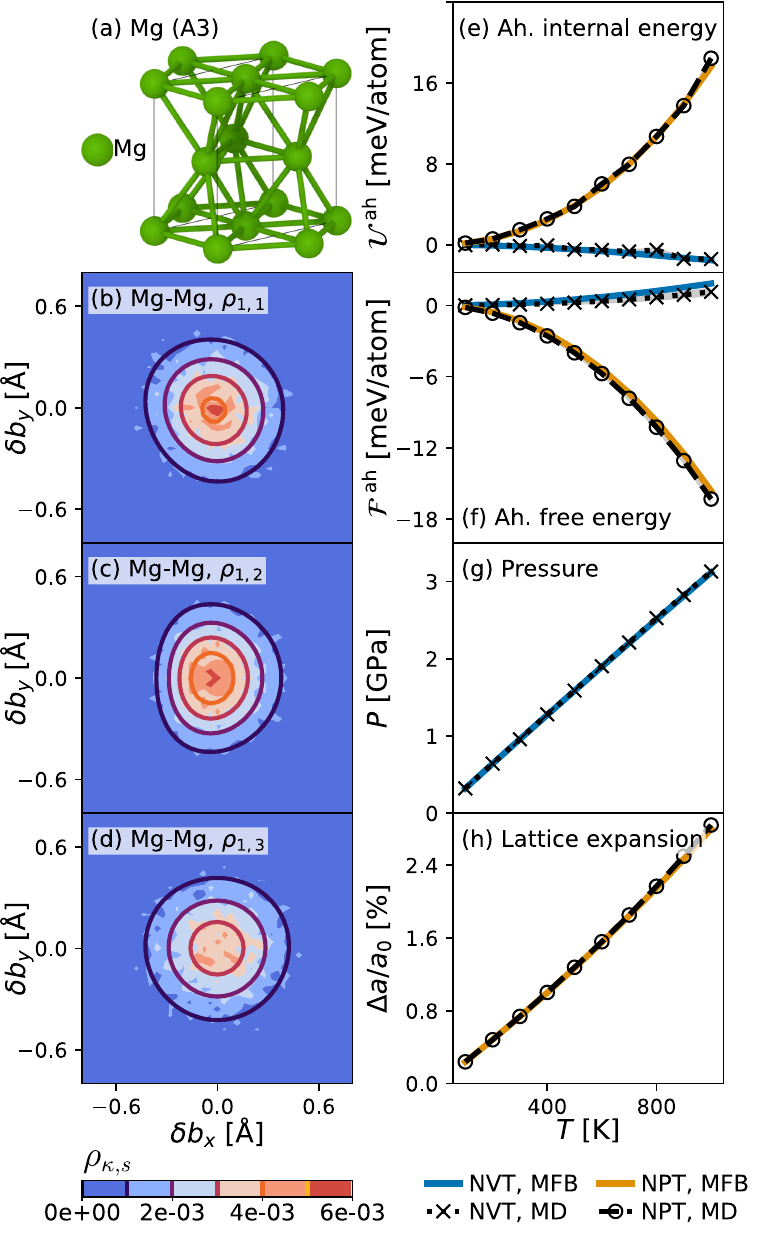}
            \caption{Bond densities and thermodynamics of Mg. (a) Unit cell. (b)--(c) Mg-Mg bond density for the two site-symmetry orbits of the first physical neighbor shell, and (d) for the second physical neighbor shell, filled contours from MD and line contours from the converged generalized mean-field bond (MFB) model, both at $1000\,\mathrm{K}$ under NVT. (e) Anharmonic internal energy, (f) anharmonic free energy, (g) pressure, and (h) lattice expansion versus temperature, comparing the generalized model against MD in the NVT and NPT ensembles.}
            \label{fig:Mg-thermo}
        \end{figure}

        As the baseline case, fcc Al has a single chemical species occupying a single Wyckoff position, with three retained nearest-neighbor shells.
        Of these, only the first carries a bonding potential, while the second and third are assigned none~\cite{Dsouza2026data}.
        The bond densities of the second and third shells in panels (c) and (d) of Fig.~\ref{fig:Al-thermo} are therefore dictated entirely by their coupling to the first shell through Eq.~\eqref{eq:predbond}, and the MFB model reproduces all three, together with the resulting thermodynamics in both ensembles.

        bcc Fe likewise has a single chemical species occupying a single Wyckoff position with three retained nearest-neighbor shells, but unlike fcc Al, both its first and second nearest-neighbor shells carry bonding potentials of their own.
        Panels (b)--(d) of Fig.~\ref{fig:Fe-thermo} show the Fe-Fe bond density for each of the three shells, where the mean-field line contours closely track the shape of the MD-sampled density for each shell, including its departure from a simple radially symmetric form at the nearest shells.
        Panels (e)--(h) show the same close agreement carried through to the ensemble-averaged thermodynamics, with the anharmonic internal and free energies, pressure, and lattice expansion predicted by the generalized mean-field bond model matching MD across the full temperature range in both ensembles.

        hcp Mg tests the reduction of two basis sites to a single representative atom.
        Its non-ideal $c/a$ ratio splits the twelve first nearest neighbors into two orbits, six in-plane and six out-of-plane, which are retained as separate shells (Sec.~\ref{sec:methods-computational}).
        As in fcc Al, only the first physical neighbor shell carries a bonding potential, but split into two shells with potentials of their own.
        Despite their nearly identical bond lengths, the two shells have bond densities of different shape as shown in panels (b) and (c) of Fig.~\ref{fig:Mg-thermo}, with one closer to axial symmetry and the other elongated, which the MFB model reproduces.
        We note that the bonding potential for the second nearest neighbors (third shell, owing to the splitting of the first) has a negative curvature at the reference volume and is therefore assigned no bonding potential.
        Similar to fcc Al, its bond density in panel (d) is shaped entirely by its coupling to the first two shells through Eq.~\eqref{eq:predbond}, and its agreement with MD is a direct test of that coupling.

        \subsection{Binary systems}
        \label{sec:results:binary}
    
\begin{figure}[b]
            \raggedright
            \includegraphics[width=1.0\columnwidth]{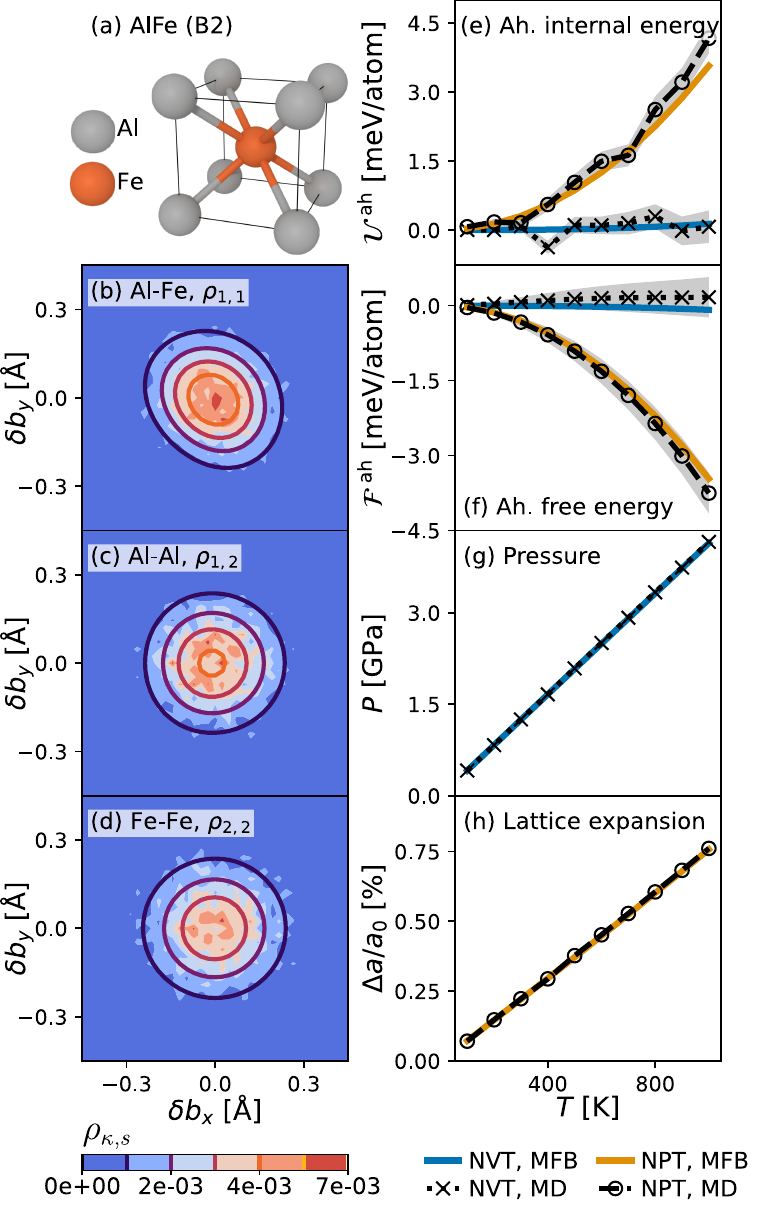}
            \caption{Bond densities and thermodynamics of AlFe. (a) Unit cell. (b) Al-Fe bond density for the shared shell $s=1$, and (c)--(d) Al-Al and Fe-Fe bond density for the second shell $s=2$ of each Wyckoff position, filled contours from MD and line contours from the converged generalized mean-field bond (MFB) model, both at $1000\,\mathrm{K}$ under NVT. (e) Anharmonic internal energy, (f) anharmonic free energy, (g) pressure, and (h) lattice expansion versus temperature, comparing the generalized model against MD in the NVT and NPT ensembles.}
            \label{fig:AlFe-thermo}
        \end{figure}

        AlFe, with its B2 structure, is the minimal case of a reciprocal pair joining two Wyckoff positions.
        Panel (b) of Fig.~\ref{fig:AlFe-thermo} shows the Al-Fe bond density of the first shell, the only shell shared between the two Wyckoff positions in our test, which by construction is identical from either end (Sec.~\ref{sec:theory-selfconsistency}).
        Panels (c) and (d) show the second shells of the two Wyckoff positions, Al-Al and Fe-Fe, each of which is its own reciprocal partner.
        The MFB model reproduces all three bond densities, including that of the shared shell.

\begin{figure}[b]
            \raggedright
            \includegraphics[width=1.0\columnwidth]{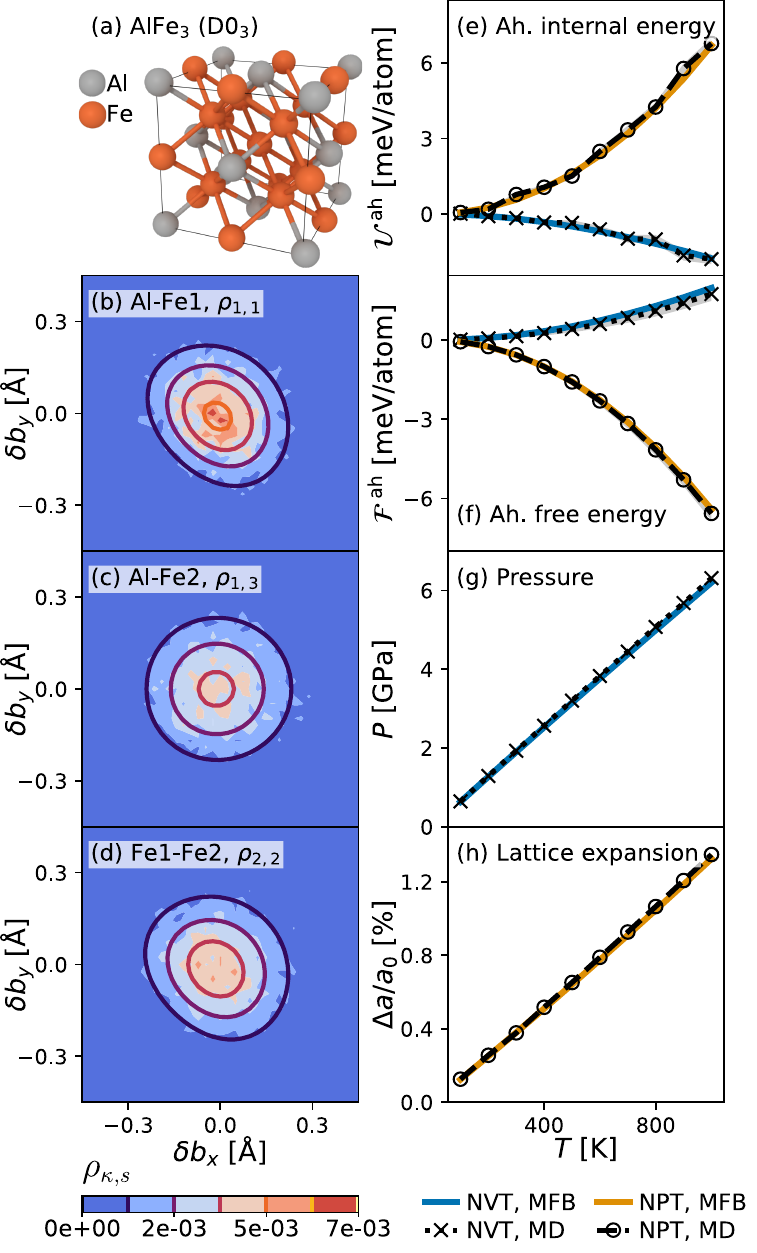}
            \caption{Bond densities and thermodynamics of AlFe$_3$. (a) Unit cell. (b) Al-Fe1 bond density for one of the two identical orbit-split shells, (c) Al-Fe2 bond density for a second, distinct shell of Al, and (d) Fe1-Fe2 bond density, connecting the two inequivalent Fe sites and sharing its bond length with (b), filled contours from MD and line contours from the converged generalized mean-field bond (MFB) model, both at $1000\,\mathrm{K}$ under NVT. (e) Anharmonic internal energy, (f) anharmonic free energy, (g) pressure, and (h) lattice expansion versus temperature, comparing the generalized model against MD in the NVT and NPT ensembles.}
            \label{fig:AlFe3-thermo}
        \end{figure}

        AlFe$_3$, with its D0$_3$ structure, adds a third Wyckoff position without a third chemical species, requiring the model to keep the two Fe sites, Fe1 and Fe2, distinct.
        Panels (b) and (c) of Fig.~\ref{fig:AlFe3-thermo} show the Al-Fe1 and Al-Fe2 bond densities, two physically distinct shells of Al, and panel (d) shows the Fe1-Fe2 bond density, which connects the two inequivalent Fe sites.
        These shared bonds couple all three Wyckoff positions, with each described identically from either end at the single effective temperature of the crystal.
        The MFB model accurately reproduces all three bond densities as well as the resulting thermodynamics.

        \subsection{Ternary system}
        \label{sec:results:ternary}
    
\begin{figure}[b]
            \raggedright
            \includegraphics[width=1.0\columnwidth]{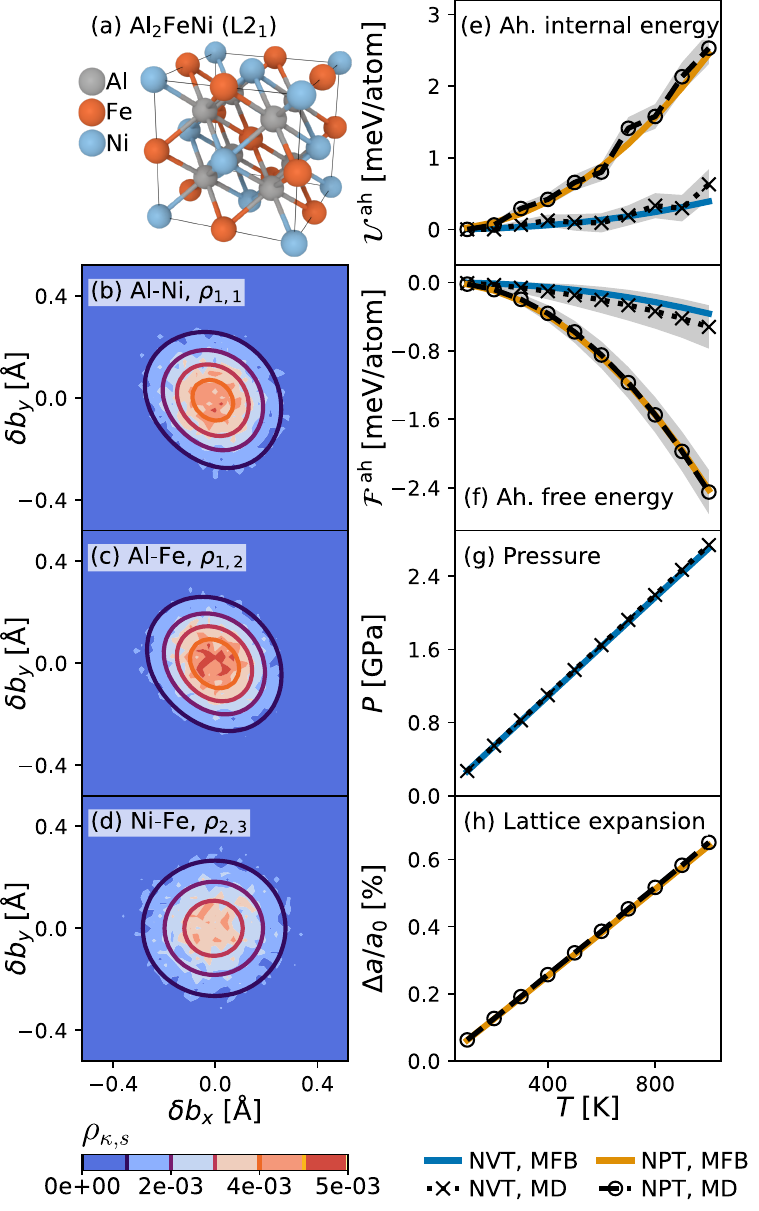}
            \caption{Bond densities and thermodynamics of Al$_2$FeNi. (a) Unit cell. (b) Al-Ni bond density, (c) Al-Fe bond density, both orbit-split on the Fe and Ni sides but not on the side of Al, and (d) Ni-Fe bond density, a separate physical distance, filled contours from MD and line contours from the converged generalized mean-field bond (MFB) model, both at $1000\,\mathrm{K}$ under NVT. (e) Anharmonic internal energy, (f) anharmonic free energy, (g) pressure, and (h) lattice expansion versus temperature, comparing the generalized model against MD in the NVT and NPT ensembles.}
            \label{fig:Al2FeNi-thermo}
        \end{figure}

        The Heusler alloy Al$_2$FeNi, with its L2$_1$ structure, combines three chemical species on three Wyckoff positions.
        Al sits at the same distance from both Fe and Ni, giving two chemically distinct first-neighbor bonds of equal length, whose bond densities are shown in panels (b) Al-Ni and (c) Al-Fe of Fig.~\ref{fig:Al2FeNi-thermo}.
        Panel (d) shows the Ni-Fe bond density, a second nearest-neighbor bond that connects the two minority Wyckoff positions directly.
        All three bonds are shared between different Wyckoff positions, coupling all three, as in AlFe$_3$, and the MFB model reproduces all three bond densities and the resulting thermodynamics.

    \section{Discussion}
    \label{sec:discussion}
    
    The results presented in Sec.~\ref{sec:results} establish that the generalized mean-field bond model reproduces finite-temperature thermodynamics across structures of increasing complexity.
    Two questions remain that determine its applicability with DFT and foundation models, namely its computational cost, and its accuracy relative to a target potential energy surface.
    The generalized model retains the convergence speed of the original formulation, such that once the bonding potentials and covariances are obtained, the self-consistent solution requires $0.7$--$1.6$ core seconds per temperature in the NVT ensemble for crystals with up to three Wyckoff positions and three retained shells, and $3.9$--$6.5$ core seconds in the NPT ensemble, where the NVT conditions are re-solved at every trial strain.
    
    More importantly for applications with DFT and foundation models, the generalized model requires no force evaluations at finite temperature.
    Force evaluations enter only through the parametrization of the bonding potentials, while the covariances follow from the force constants of the bond lattice, which are obtained from the bonding potentials themselves without further force evaluations.
    The number of force evaluations is therefore independent of the number of temperatures of interest, and, for bonding potentials calibrated jointly across a range of volumes as in Sec.~\ref{sec:methods-parametrization}, independent of the volumes within that range, matching the static calculations required by the QHA.
    If a single bonding potential per shell cannot capture how the interactions of the target change with volume, the bonding potentials must instead be parametrized at each volume, and the number of force evaluations then grows with the number of volumes, as in the QHA.
    
    Since the fully correlated mean-field approximation recovers the exact harmonic single bond density (Sec.~\ref{sec:theory-covariance}), and reproduces MD with identical bonding potentials across the structures considered here, the accuracy of the model with respect to a target potential energy surface is set predominantly by how well the bonding potentials represent it.
    The Morse-type potentials used in this work were calibrated only to the curvature of the reference interatomic potentials at a few volumes, and were chosen to isolate the error of the generalized model rather than to reproduce the anharmonicity of the potentials they were parametrized from.
    Predicting accurate finite-temperature properties at \emph{ab initio} accuracy therefore reduces to parametrizing bonding potentials that capture the anharmonicity of the target potential energy surface.
    The same holds for the number of shells that carry bonding potentials and their relative contributions, which reflect the range of the bonding potentials rather than a property of the model, as the model accommodates any number of shells, including shells without a bonding potential of their own, whose densities follow from their coupling to the other shells, as in fcc Al and hcp Mg.
    
    The local anharmonic approximation of Glensk \textit{et al.}~\cite{Glensk2015} provides such a parametrization from static calculations alone, by displacing an atom along the longitudinal and two transverse directions to each bond and fitting the bonding potential from the resulting forces.
    The procedure extends directly to the present generalization, with one bonding potential fitted per shell and Wyckoff position, and with the displacements of both ends of a reciprocal pair fitted jointly to satisfy the reciprocity of Eq.~\eqref{eq:reciprocity}.
    As the displacements sample the anharmonic regime directly over amplitudes far larger than those captured by the force constants, the resulting potentials capture the anharmonicity of the target potential energy surface, rather than only its curvature.
    The required static calculations number in the tens to hundreds per structure, placing anharmonic thermodynamics within reach of DFT directly, at a cost comparable to that of the QHA.
    Alternatively, the bonding potentials can be parametrized from a foundation model, with the resulting free energy corrected to \emph{ab initio} accuracy by upsampled TI~\cite{Grabowski2009}, running DFT calculations on a small number of configurations sampled from the bond lattice, whose own free energy the model provides without sampling.
    
    In their current form, bonding potentials parametrized from the local anharmonic approximation are additively separable along the longitudinal and two transverse directions of each bond, and therefore neglect any coupling between these components~\cite{Glensk2015, Dsouza2024}.
    More fundamentally, a bonding potential that depends on a single bond folds the many-body contributions of the target potential energy surface, such as those arising from bond angles, into pairwise terms only partially.
    A natural extension is a minimal expansion in two- and three-body terms in a symmetry-adapted basis, as in the atomic cluster expansion~\cite{Drautz2019}, in which the symmetry of each shell and Wyckoff position restricts the allowed terms, and the truncation order fixes the number of parameters.
    Since Eq.~\eqref{eq:predbond} expresses every bond in the crystal as a function of the reference bond, three-body terms enter the effective potential through the same predicted bonds as two-body terms, leaving the fully correlated mean-field approximation itself unchanged.
    
    A key consideration, however, is that the cost of the model is set by the bonding potentials rather than by the mean-field approximation itself.
    The effective potential of Eq.~\eqref{eq:Veff} evaluates every bonding potential retained in the sum at every point of the bond grid, such that the cost of a single evaluation is multiplied by the number of bonds in the bond-centered cluster, which grows with the number of retained shells and Wyckoff positions.
    The self-consistent solution of Sec.~\ref{sec:theory-selfconsistency}, in contrast, adds only one-dimensional root searches for a single effective temperature and one Lagrange multiplier per distinct shell, as shells related by symmetry or reciprocity share a single density.
    A cluster expansion therefore preserves the sampling-free efficiency of the generalized model only if it remains compact, with the truncation order set by the trade-off between the accuracy of the bonding potentials and the cost of evaluating the effective potential.
    The construction and parametrization of such expansions will be addressed in future work.
    
    The present formulation retains a few limitations that bound its scope.
    As the coupling between the bonds is harmonic, the anharmonic correlations between the bonds are neglected, an approximation that can become more severe with increasing temperature and anharmonicity.
    Moreover, the covariances are evaluated once from the force constants at $\epsilon=0$.
    Although the normalization of Eq.~\eqref{eq:alpha} removes any uniform rescaling of the force constants, the shells can stiffen at different rates under strain, and stretched bonds can acquire a transverse stiffness absent at $\epsilon=0$, such that the relative coupling between bonds varies weakly with strain.
    Since the force constants of the bond lattice at any strain follow from the bonding potentials themselves, this approximation can be removed without additional force evaluations, at the cost of re-evaluating the covariances at every trial strain.
    Another limitation is that while the isotropic strain is exact for cubic crystals, it fixes the cell shape of lower-symmetry crystals such as hcp Mg.
    A complete treatment of such crystals would require replacing the isotropic strain with a deformation gradient acting on every equilibrium bond, and solving for the cell shape alongside the volume from the full stress tensor in the NPT ensemble.
    Finally, the model is classical and assumes a fixed bond topology, such that it excludes zero-point and quantum vibrational effects, which can be significant below the Debye temperature for light elements~\cite{Herrero2014, Markland2018}, as well as diffusion and melting.
    Quantum effects, however, can be approximated by evaluating the model at remapped classical temperatures~\cite{Dsouza2022}, which we will also address in future work.
    
    \section{Conclusion}
    \label{sec:conclusion}
    
    We have generalized the mean-field bond model to arbitrary neighbor shells in crystals with multiple sublattices and chemical species.
    Every atom of the crystal is reduced by symmetry to one representative atom per Wyckoff position, and the fluctuations of every bond are coupled to those of the reference bond of each shell through normalized harmonic covariances obtained from the force constants of the relaxed crystal, with the bonds coupled to each reference bond taken around both of its ends.
    These covariances are independent of temperature and atomic mass, satisfy material compatibility identically, and recover the exact harmonic single bond density, such that the approximation is exact for harmonic bonding potentials and introduces error only through the harmonic coupling between bonds, the truncation of the coupling, and the strain dependence of the covariances.
    Thermodynamic self-consistency is enforced through the volume constraint of every shell and a single virial temperature constraint for the crystal, such that every shared bond is described identically from either end.
    
    We validated the generalized model against MD with identical Morse-type bonding potentials for six crystal structures of increasing complexity, namely fcc Al, bcc Fe, and hcp Mg, the B2 and D0$_3$ alloys AlFe and AlFe$_3$, and the L2$_1$ Heusler alloy Al$_2$FeNi.
    In both the isochoric and isobaric ensembles, the model reproduces the bond densities, anharmonic internal and free energies, pressures, and thermal expansion of MD, to within its statistical uncertainty, at a cost of a few core seconds per temperature.
    
    Since the model requires no force evaluations at finite temperature, its computational cost with DFT or a foundation model is set by the static calculations needed to parametrize the bonding potentials, independent of the number of temperatures of interest, while its accuracy is set by how well these potentials represent the target potential energy surface.
    Bonding potentials parametrized within the local anharmonic approximation, and extended to compact many-body expansions that capture the coupling neglected by separable forms, therefore offer a route to the finite-temperature thermodynamics of complex crystals at \emph{ab initio} accuracy without finite-temperature force evaluations.

    \section*{Code and data availability}
    \label{sec:code}
    
    The implementation of the generalized mean-field bond model is available at \url{https://github.com/raynol-dsouza/mean-field-bond-model}, and version 1.0.0, used for this work, is archived at Zenodo~\cite{Dsouza2026code}, together with the notebooks required to reproduce the figures and tables presented in this paper.
    The relaxed structures, the reference interatomic potentials, the fitted parameters of the Morse-type bonding potentials, the force constants, the MD data and the model results are archived at Zenodo~\cite{Dsouza2026data}, from which the notebooks download them.
    
    \begin{acknowledgments}
        R.D. is grateful for support from the Max Planck Society and the CNRS through the Salto Exchange Programme, which enabled a research stay at CINaM UMR 7325, Aix-Marseille Universit\'e, where part of this research was conducted.
        R.D. also thanks Liam Huber and Marvin Poul for stimulating discussions and valuable feedback.
        T.D.S. gratefully acknowledges support from a MICDE Catalyst Grant, University of Michigan and ANR grant ANR-23-CE46-0006-1.
    \end{acknowledgments}

    \appendix
    \section{Mass cancellation in the displacement covariance}
    \label{app:mass-cancellation}
    
    The conventional phonon dynamical matrix is mass-weighted
    \begin{equation}
        \bar{\mathbf D}(\mathbf q) \equiv \mathbf M^{-1/2}\,\mathbf D(\mathbf q)\,\mathbf M^{-1/2},
        \label{eq:massweighted}
    \end{equation}
    where $\mathbf M$ is the $3B\times3B$ block-diagonal mass matrix with blocks $\mathbf M_{\mu\mu} = m_\mu\mathbb I$, and $m_\mu$ is the mass of the atoms on basis site $\mu$.
    Equipartition then applies to the mass-weighted displacements $\delta\bar{\mathbf R}_i \equiv \sqrt{m_{\mu(i)}}\,\delta\mathbf R_i$, whose covariance can be written, as in Eq.~\eqref{eq:greensfunction}, as
    \begin{equation}
        \langle \delta\bar{\mathbf R}_i \otimes \delta\bar{\mathbf R}_j \rangle
        = \frac{k_{\mathrm B}T}{N_c}\sum_{\mathbf q} \bar{\mathbf D}^{+}_{\mu(i)\mu(j)}(\mathbf q)\,e^{i\mathbf q\cdot(\mathbf R_i-\mathbf R_j)}.
        \label{eq:massweightedcov}
    \end{equation}
    Since the mass weighting in Eq.~\eqref{eq:massweighted} is diagonal in the basis sites, it inverts block-wise as
    \begin{equation}
        \bar{\mathbf D}^{+}_{\mu\mu'}(\mathbf q) = \sqrt{m_\mu m_{\mu'}}\;\mathbf D^{+}_{\mu\mu'}(\mathbf q),
        \label{eq:massinverse}
    \end{equation}
    while, by definition, 
    \begin{equation}
        \langle \delta\bar{\mathbf R}_i \otimes \delta\bar{\mathbf R}_j \rangle
        \equiv \sqrt{m_{\mu(i)}m_{\mu(j)}}\,\langle \delta\mathbf R_i \otimes \delta\mathbf R_j \rangle,
        \label{eq:massweightedrelation}
    \end{equation}
    Substituting Eqs.~\eqref{eq:massinverse} and~\eqref{eq:massweightedrelation} into Eq.~\eqref{eq:massweightedcov}, the factor $\sqrt{m_{\mu(i)}m_{\mu(j)}}$ appears on both sides and cancels, giving Eq.~\eqref{eq:posvariance} for the unweighted displacements, independent of the masses.
    At $\mathbf q = 0$, where both matrices are singular, Eq.~\eqref{eq:massinverse} holds only up to a uniform translation of the crystal, which leaves every bond deviation unchanged.
    
    \bibliographystyle{apsrev4-2}
    \bibliography{references}

\begin{thebibliography}{60}%
\makeatletter
\providecommand \@ifxundefined [1]{%
 \@ifx{#1\undefined}
}%
\providecommand \@ifnum [1]{%
 \ifnum #1\expandafter \@firstoftwo
 \else \expandafter \@secondoftwo
 \fi
}%
\providecommand \@ifx [1]{%
 \ifx #1\expandafter \@firstoftwo
 \else \expandafter \@secondoftwo
 \fi
}%
\providecommand \natexlab [1]{#1}%
\providecommand \enquote  [1]{``#1''}%
\providecommand \bibnamefont  [1]{#1}%
\providecommand \bibfnamefont [1]{#1}%
\providecommand \citenamefont [1]{#1}%
\providecommand \href@noop [0]{\@secondoftwo}%
\providecommand \href [0]{\begingroup \@sanitize@url \@href}%
\providecommand \@href[1]{\@@startlink{#1}\@@href}%
\providecommand \@@href[1]{\endgroup#1\@@endlink}%
\providecommand \@sanitize@url [0]{\catcode `\\12\catcode `\$12\catcode `\&12\catcode `\#12\catcode `\^12\catcode `\_12\catcode `\%12\relax}%
\providecommand \@@startlink[1]{}%
\providecommand \@@endlink[0]{}%
\providecommand \url  [0]{\begingroup\@sanitize@url \@url }%
\providecommand \@url [1]{\endgroup\@href {#1}{\urlprefix }}%
\providecommand \urlprefix  [0]{URL }%
\providecommand \Eprint [0]{\href }%
\providecommand \doibase [0]{https://doi.org/}%
\providecommand \selectlanguage [0]{\@gobble}%
\providecommand \bibinfo  [0]{\@secondoftwo}%
\providecommand \bibfield  [0]{\@secondoftwo}%
\providecommand \translation [1]{[#1]}%
\providecommand \BibitemOpen [0]{}%
\providecommand \bibitemStop [0]{}%
\providecommand \bibitemNoStop [0]{.\EOS\space}%
\providecommand \EOS [0]{\spacefactor3000\relax}%
\providecommand \BibitemShut  [1]{\csname bibitem#1\endcsname}%
\let\auto@bib@innerbib\@empty
\bibitem [{\citenamefont {Wallace}(1972)}]{Wallace1972}%
  \BibitemOpen
  \bibfield  {author} {\bibinfo {author} {\bibfnamefont {D.~C.}\ \bibnamefont {Wallace}},\ }\href@noop {} {\emph {\bibinfo {title} {Thermodynamics of Crystals}}}\ (\bibinfo  {publisher} {John Wiley \& Sons},\ \bibinfo {address} {New York},\ \bibinfo {year} {1972})\BibitemShut {NoStop}%
\bibitem [{\citenamefont {Frenkel}\ and\ \citenamefont {Smit}(2002)}]{Frenkel2002}%
  \BibitemOpen
  \bibfield  {author} {\bibinfo {author} {\bibfnamefont {D.}~\bibnamefont {Frenkel}}\ and\ \bibinfo {author} {\bibfnamefont {B.}~\bibnamefont {Smit}},\ }\href@noop {} {\emph {\bibinfo {title} {Understanding Molecular Simulation: From Algorithms to Applications}}},\ \bibinfo {edition} {2nd}\ ed.\ (\bibinfo  {publisher} {Academic Press},\ \bibinfo {address} {San Diego},\ \bibinfo {year} {2002})\BibitemShut {NoStop}%
\bibitem [{\citenamefont {Hohenberg}\ and\ \citenamefont {Kohn}(1964)}]{Hohenberg1964}%
  \BibitemOpen
  \bibfield  {author} {\bibinfo {author} {\bibfnamefont {P.}~\bibnamefont {Hohenberg}}\ and\ \bibinfo {author} {\bibfnamefont {W.}~\bibnamefont {Kohn}},\ }\href {https://doi.org/10.1103/PhysRev.136.B864} {\bibfield  {journal} {\bibinfo  {journal} {Phys. Rev.}\ }\textbf {\bibinfo {volume} {136}},\ \bibinfo {pages} {B864} (\bibinfo {year} {1964})}\BibitemShut {NoStop}%
\bibitem [{\citenamefont {Kohn}\ and\ \citenamefont {Sham}(1965)}]{Kohn1965}%
  \BibitemOpen
  \bibfield  {author} {\bibinfo {author} {\bibfnamefont {W.}~\bibnamefont {Kohn}}\ and\ \bibinfo {author} {\bibfnamefont {L.~J.}\ \bibnamefont {Sham}},\ }\href {https://doi.org/10.1103/PhysRev.140.A1133} {\bibfield  {journal} {\bibinfo  {journal} {Phys. Rev.}\ }\textbf {\bibinfo {volume} {140}},\ \bibinfo {pages} {A1133} (\bibinfo {year} {1965})}\BibitemShut {NoStop}%
\bibitem [{\citenamefont {Baroni}\ \emph {et~al.}(2001)\citenamefont {Baroni}, \citenamefont {de~Gironcoli}, \citenamefont {Dal~Corso},\ and\ \citenamefont {Giannozzi}}]{Baroni2001}%
  \BibitemOpen
  \bibfield  {author} {\bibinfo {author} {\bibfnamefont {S.}~\bibnamefont {Baroni}}, \bibinfo {author} {\bibfnamefont {S.}~\bibnamefont {de~Gironcoli}}, \bibinfo {author} {\bibfnamefont {A.}~\bibnamefont {Dal~Corso}},\ and\ \bibinfo {author} {\bibfnamefont {P.}~\bibnamefont {Giannozzi}},\ }\href {https://doi.org/10.1103/RevModPhys.73.515} {\bibfield  {journal} {\bibinfo  {journal} {Rev. Mod. Phys.}\ }\textbf {\bibinfo {volume} {73}},\ \bibinfo {pages} {515} (\bibinfo {year} {2001})}\BibitemShut {NoStop}%
\bibitem [{\citenamefont {Togo}(2023)}]{Togo2023}%
  \BibitemOpen
  \bibfield  {author} {\bibinfo {author} {\bibfnamefont {A.}~\bibnamefont {Togo}},\ }\href {https://doi.org/10.7566/JPSJ.92.012001} {\bibfield  {journal} {\bibinfo  {journal} {J. Phys. Soc. Jpn.}\ }\textbf {\bibinfo {volume} {92}},\ \bibinfo {pages} {012001} (\bibinfo {year} {2023})}\BibitemShut {NoStop}%
\bibitem [{\citenamefont {Grabowski}\ \emph {et~al.}(2009)\citenamefont {Grabowski}, \citenamefont {Ismer}, \citenamefont {Hickel},\ and\ \citenamefont {Neugebauer}}]{Grabowski2009}%
  \BibitemOpen
  \bibfield  {author} {\bibinfo {author} {\bibfnamefont {B.}~\bibnamefont {Grabowski}}, \bibinfo {author} {\bibfnamefont {L.}~\bibnamefont {Ismer}}, \bibinfo {author} {\bibfnamefont {T.}~\bibnamefont {Hickel}},\ and\ \bibinfo {author} {\bibfnamefont {J.}~\bibnamefont {Neugebauer}},\ }\href {https://doi.org/10.1103/PhysRevB.79.134106} {\bibfield  {journal} {\bibinfo  {journal} {Phys. Rev. B}\ }\textbf {\bibinfo {volume} {79}},\ \bibinfo {pages} {134106} (\bibinfo {year} {2009})}\BibitemShut {NoStop}%
\bibitem [{\citenamefont {Glensk}\ \emph {et~al.}(2014)\citenamefont {Glensk}, \citenamefont {Grabowski}, \citenamefont {Hickel},\ and\ \citenamefont {Neugebauer}}]{Glensk2014}%
  \BibitemOpen
  \bibfield  {author} {\bibinfo {author} {\bibfnamefont {A.}~\bibnamefont {Glensk}}, \bibinfo {author} {\bibfnamefont {B.}~\bibnamefont {Grabowski}}, \bibinfo {author} {\bibfnamefont {T.}~\bibnamefont {Hickel}},\ and\ \bibinfo {author} {\bibfnamefont {J.}~\bibnamefont {Neugebauer}},\ }\href {https://doi.org/10.1103/PhysRevX.4.011018} {\bibfield  {journal} {\bibinfo  {journal} {Phys. Rev. X}\ }\textbf {\bibinfo {volume} {4}},\ \bibinfo {pages} {011018} (\bibinfo {year} {2014})}\BibitemShut {NoStop}%
\bibitem [{\citenamefont {Glensk}\ \emph {et~al.}(2015)\citenamefont {Glensk}, \citenamefont {Grabowski}, \citenamefont {Hickel},\ and\ \citenamefont {Neugebauer}}]{Glensk2015}%
  \BibitemOpen
  \bibfield  {author} {\bibinfo {author} {\bibfnamefont {A.}~\bibnamefont {Glensk}}, \bibinfo {author} {\bibfnamefont {B.}~\bibnamefont {Grabowski}}, \bibinfo {author} {\bibfnamefont {T.}~\bibnamefont {Hickel}},\ and\ \bibinfo {author} {\bibfnamefont {J.}~\bibnamefont {Neugebauer}},\ }\href {https://doi.org/10.1103/PhysRevLett.114.195901} {\bibfield  {journal} {\bibinfo  {journal} {Phys. Rev. Lett.}\ }\textbf {\bibinfo {volume} {114}},\ \bibinfo {pages} {195901} (\bibinfo {year} {2015})}\BibitemShut {NoStop}%
\bibitem [{\citenamefont {Kirkwood}(1935)}]{Kirkwood1935}%
  \BibitemOpen
  \bibfield  {author} {\bibinfo {author} {\bibfnamefont {J.~G.}\ \bibnamefont {Kirkwood}},\ }\href {https://doi.org/10.1063/1.1749657} {\bibfield  {journal} {\bibinfo  {journal} {J. Chem. Phys.}\ }\textbf {\bibinfo {volume} {3}},\ \bibinfo {pages} {300} (\bibinfo {year} {1935})}\BibitemShut {NoStop}%
\bibitem [{\citenamefont {Frenkel}\ and\ \citenamefont {Ladd}(1984)}]{Frenkel1984}%
  \BibitemOpen
  \bibfield  {author} {\bibinfo {author} {\bibfnamefont {D.}~\bibnamefont {Frenkel}}\ and\ \bibinfo {author} {\bibfnamefont {A.~J.~C.}\ \bibnamefont {Ladd}},\ }\href {https://doi.org/10.1063/1.448024} {\bibfield  {journal} {\bibinfo  {journal} {J. Chem. Phys.}\ }\textbf {\bibinfo {volume} {81}},\ \bibinfo {pages} {3188} (\bibinfo {year} {1984})}\BibitemShut {NoStop}%
\bibitem [{\citenamefont {Zwanzig}(1954)}]{Zwanzig1954}%
  \BibitemOpen
  \bibfield  {author} {\bibinfo {author} {\bibfnamefont {R.~W.}\ \bibnamefont {Zwanzig}},\ }\href {https://doi.org/10.1063/1.1740409} {\bibfield  {journal} {\bibinfo  {journal} {J. Chem. Phys.}\ }\textbf {\bibinfo {volume} {22}},\ \bibinfo {pages} {1420} (\bibinfo {year} {1954})}\BibitemShut {NoStop}%
\bibitem [{\citenamefont {Duff}\ \emph {et~al.}(2015)\citenamefont {Duff}, \citenamefont {Davey}, \citenamefont {Korbmacher}, \citenamefont {Glensk}, \citenamefont {Grabowski}, \citenamefont {Neugebauer},\ and\ \citenamefont {Finnis}}]{Duff2015}%
  \BibitemOpen
  \bibfield  {author} {\bibinfo {author} {\bibfnamefont {A.~I.}\ \bibnamefont {Duff}}, \bibinfo {author} {\bibfnamefont {T.}~\bibnamefont {Davey}}, \bibinfo {author} {\bibfnamefont {D.}~\bibnamefont {Korbmacher}}, \bibinfo {author} {\bibfnamefont {A.}~\bibnamefont {Glensk}}, \bibinfo {author} {\bibfnamefont {B.}~\bibnamefont {Grabowski}}, \bibinfo {author} {\bibfnamefont {J.}~\bibnamefont {Neugebauer}},\ and\ \bibinfo {author} {\bibfnamefont {M.~W.}\ \bibnamefont {Finnis}},\ }\href {https://doi.org/10.1103/PhysRevB.91.214311} {\bibfield  {journal} {\bibinfo  {journal} {Phys. Rev. B}\ }\textbf {\bibinfo {volume} {91}},\ \bibinfo {pages} {214311} (\bibinfo {year} {2015})}\BibitemShut {NoStop}%
\bibitem [{\citenamefont {Grabowski}\ \emph {et~al.}(2019)\citenamefont {Grabowski}, \citenamefont {Ikeda}, \citenamefont {Srinivasan}, \citenamefont {K\"ormann}, \citenamefont {Freysoldt}, \citenamefont {Duff}, \citenamefont {Shapeev},\ and\ \citenamefont {Neugebauer}}]{Grabowski2019}%
  \BibitemOpen
  \bibfield  {author} {\bibinfo {author} {\bibfnamefont {B.}~\bibnamefont {Grabowski}}, \bibinfo {author} {\bibfnamefont {Y.}~\bibnamefont {Ikeda}}, \bibinfo {author} {\bibfnamefont {P.}~\bibnamefont {Srinivasan}}, \bibinfo {author} {\bibfnamefont {F.}~\bibnamefont {K\"ormann}}, \bibinfo {author} {\bibfnamefont {C.}~\bibnamefont {Freysoldt}}, \bibinfo {author} {\bibfnamefont {A.~I.}\ \bibnamefont {Duff}}, \bibinfo {author} {\bibfnamefont {A.}~\bibnamefont {Shapeev}},\ and\ \bibinfo {author} {\bibfnamefont {J.}~\bibnamefont {Neugebauer}},\ }\href {https://doi.org/10.1038/s41524-019-0218-8} {\bibfield  {journal} {\bibinfo  {journal} {npj Comput. Mater.}\ }\textbf {\bibinfo {volume} {5}},\ \bibinfo {pages} {80} (\bibinfo {year} {2019})}\BibitemShut {NoStop}%
\bibitem [{\citenamefont {Jung}\ \emph {et~al.}(2023)\citenamefont {Jung}, \citenamefont {Srinivasan}, \citenamefont {Forslund},\ and\ \citenamefont {Grabowski}}]{Jung2023}%
  \BibitemOpen
  \bibfield  {author} {\bibinfo {author} {\bibfnamefont {J.~H.}\ \bibnamefont {Jung}}, \bibinfo {author} {\bibfnamefont {P.}~\bibnamefont {Srinivasan}}, \bibinfo {author} {\bibfnamefont {A.}~\bibnamefont {Forslund}},\ and\ \bibinfo {author} {\bibfnamefont {B.}~\bibnamefont {Grabowski}},\ }\href {https://doi.org/10.1038/s41524-022-00956-8} {\bibfield  {journal} {\bibinfo  {journal} {npj Comput. Mater.}\ }\textbf {\bibinfo {volume} {9}},\ \bibinfo {pages} {3} (\bibinfo {year} {2023})}\BibitemShut {NoStop}%
\bibitem [{\citenamefont {Hellman}\ \emph {et~al.}(2011)\citenamefont {Hellman}, \citenamefont {Abrikosov},\ and\ \citenamefont {Simak}}]{Hellman2011}%
  \BibitemOpen
  \bibfield  {author} {\bibinfo {author} {\bibfnamefont {O.}~\bibnamefont {Hellman}}, \bibinfo {author} {\bibfnamefont {I.~A.}\ \bibnamefont {Abrikosov}},\ and\ \bibinfo {author} {\bibfnamefont {S.~I.}\ \bibnamefont {Simak}},\ }\href {https://doi.org/10.1103/PhysRevB.84.180301} {\bibfield  {journal} {\bibinfo  {journal} {Phys. Rev. B}\ }\textbf {\bibinfo {volume} {84}},\ \bibinfo {pages} {180301(R)} (\bibinfo {year} {2011})}\BibitemShut {NoStop}%
\bibitem [{\citenamefont {Hellman}\ \emph {et~al.}(2013)\citenamefont {Hellman}, \citenamefont {Steneteg}, \citenamefont {Abrikosov},\ and\ \citenamefont {Simak}}]{Hellman2013}%
  \BibitemOpen
  \bibfield  {author} {\bibinfo {author} {\bibfnamefont {O.}~\bibnamefont {Hellman}}, \bibinfo {author} {\bibfnamefont {P.}~\bibnamefont {Steneteg}}, \bibinfo {author} {\bibfnamefont {I.~A.}\ \bibnamefont {Abrikosov}},\ and\ \bibinfo {author} {\bibfnamefont {S.~I.}\ \bibnamefont {Simak}},\ }\href {https://doi.org/10.1103/PhysRevB.87.104111} {\bibfield  {journal} {\bibinfo  {journal} {Phys. Rev. B}\ }\textbf {\bibinfo {volume} {87}},\ \bibinfo {pages} {104111} (\bibinfo {year} {2013})}\BibitemShut {NoStop}%
\bibitem [{\citenamefont {Shulumba}\ \emph {et~al.}(2017)\citenamefont {Shulumba}, \citenamefont {Hellman},\ and\ \citenamefont {Minnich}}]{Shulumba2017}%
  \BibitemOpen
  \bibfield  {author} {\bibinfo {author} {\bibfnamefont {N.}~\bibnamefont {Shulumba}}, \bibinfo {author} {\bibfnamefont {O.}~\bibnamefont {Hellman}},\ and\ \bibinfo {author} {\bibfnamefont {A.~J.}\ \bibnamefont {Minnich}},\ }\href {https://doi.org/10.1103/PhysRevB.95.014302} {\bibfield  {journal} {\bibinfo  {journal} {Phys. Rev. B}\ }\textbf {\bibinfo {volume} {95}},\ \bibinfo {pages} {014302} (\bibinfo {year} {2017})}\BibitemShut {NoStop}%
\bibitem [{\citenamefont {Knoop}\ \emph {et~al.}(2024)\citenamefont {Knoop}, \citenamefont {Shulumba}, \citenamefont {Castellano}, \citenamefont {Batista}, \citenamefont {Farris}, \citenamefont {Verstraete}, \citenamefont {Heine}, \citenamefont {Broido}, \citenamefont {Kim}, \citenamefont {Klarbring}, \citenamefont {Abrikosov}, \citenamefont {Simak},\ and\ \citenamefont {Hellman}}]{Knoop2024}%
  \BibitemOpen
  \bibfield  {author} {\bibinfo {author} {\bibfnamefont {F.}~\bibnamefont {Knoop}}, \bibinfo {author} {\bibfnamefont {N.}~\bibnamefont {Shulumba}}, \bibinfo {author} {\bibfnamefont {A.}~\bibnamefont {Castellano}}, \bibinfo {author} {\bibfnamefont {J.~P.~A.}\ \bibnamefont {Batista}}, \bibinfo {author} {\bibfnamefont {R.}~\bibnamefont {Farris}}, \bibinfo {author} {\bibfnamefont {M.~J.}\ \bibnamefont {Verstraete}}, \bibinfo {author} {\bibfnamefont {M.}~\bibnamefont {Heine}}, \bibinfo {author} {\bibfnamefont {D.}~\bibnamefont {Broido}}, \bibinfo {author} {\bibfnamefont {D.~S.}\ \bibnamefont {Kim}}, \bibinfo {author} {\bibfnamefont {J.}~\bibnamefont {Klarbring}}, \bibinfo {author} {\bibfnamefont {I.~A.}\ \bibnamefont {Abrikosov}}, \bibinfo {author} {\bibfnamefont {S.~I.}\ \bibnamefont {Simak}},\ and\ \bibinfo {author} {\bibfnamefont {O.}~\bibnamefont {Hellman}},\ }\href {https://doi.org/10.21105/joss.06150} {\bibfield  {journal} {\bibinfo  {journal} {J. Open Source Softw.}\ }\textbf {\bibinfo {volume} {9}},\
  \bibinfo {pages} {6150} (\bibinfo {year} {2024})}\BibitemShut {NoStop}%
\bibitem [{\citenamefont {Errea}\ \emph {et~al.}(2014)\citenamefont {Errea}, \citenamefont {Calandra},\ and\ \citenamefont {Mauri}}]{Errea2014}%
  \BibitemOpen
  \bibfield  {author} {\bibinfo {author} {\bibfnamefont {I.}~\bibnamefont {Errea}}, \bibinfo {author} {\bibfnamefont {M.}~\bibnamefont {Calandra}},\ and\ \bibinfo {author} {\bibfnamefont {F.}~\bibnamefont {Mauri}},\ }\href {https://doi.org/10.1103/PhysRevB.89.064302} {\bibfield  {journal} {\bibinfo  {journal} {Phys. Rev. B}\ }\textbf {\bibinfo {volume} {89}},\ \bibinfo {pages} {064302} (\bibinfo {year} {2014})}\BibitemShut {NoStop}%
\bibitem [{\citenamefont {Bianco}\ \emph {et~al.}(2017)\citenamefont {Bianco}, \citenamefont {Errea}, \citenamefont {Paulatto}, \citenamefont {Calandra},\ and\ \citenamefont {Mauri}}]{Bianco2017}%
  \BibitemOpen
  \bibfield  {author} {\bibinfo {author} {\bibfnamefont {R.}~\bibnamefont {Bianco}}, \bibinfo {author} {\bibfnamefont {I.}~\bibnamefont {Errea}}, \bibinfo {author} {\bibfnamefont {L.}~\bibnamefont {Paulatto}}, \bibinfo {author} {\bibfnamefont {M.}~\bibnamefont {Calandra}},\ and\ \bibinfo {author} {\bibfnamefont {F.}~\bibnamefont {Mauri}},\ }\href {https://doi.org/10.1103/PhysRevB.96.014111} {\bibfield  {journal} {\bibinfo  {journal} {Phys. Rev. B}\ }\textbf {\bibinfo {volume} {96}},\ \bibinfo {pages} {014111} (\bibinfo {year} {2017})}\BibitemShut {NoStop}%
\bibitem [{\citenamefont {Monacelli}\ \emph {et~al.}(2021)\citenamefont {Monacelli}, \citenamefont {Bianco}, \citenamefont {Cherubini}, \citenamefont {Calandra}, \citenamefont {Errea},\ and\ \citenamefont {Mauri}}]{Monacelli2021}%
  \BibitemOpen
  \bibfield  {author} {\bibinfo {author} {\bibfnamefont {L.}~\bibnamefont {Monacelli}}, \bibinfo {author} {\bibfnamefont {R.}~\bibnamefont {Bianco}}, \bibinfo {author} {\bibfnamefont {M.}~\bibnamefont {Cherubini}}, \bibinfo {author} {\bibfnamefont {M.}~\bibnamefont {Calandra}}, \bibinfo {author} {\bibfnamefont {I.}~\bibnamefont {Errea}},\ and\ \bibinfo {author} {\bibfnamefont {F.}~\bibnamefont {Mauri}},\ }\href {https://doi.org/10.1088/1361-648X/ac066b} {\bibfield  {journal} {\bibinfo  {journal} {J. Phys.: Condens. Matter}\ }\textbf {\bibinfo {volume} {33}},\ \bibinfo {pages} {363001} (\bibinfo {year} {2021})}\BibitemShut {NoStop}%
\bibitem [{\citenamefont {Behler}\ and\ \citenamefont {Parrinello}(2007)}]{Behler2007}%
  \BibitemOpen
  \bibfield  {author} {\bibinfo {author} {\bibfnamefont {J.}~\bibnamefont {Behler}}\ and\ \bibinfo {author} {\bibfnamefont {M.}~\bibnamefont {Parrinello}},\ }\href {https://doi.org/10.1103/PhysRevLett.98.146401} {\bibfield  {journal} {\bibinfo  {journal} {Phys. Rev. Lett.}\ }\textbf {\bibinfo {volume} {98}},\ \bibinfo {pages} {146401} (\bibinfo {year} {2007})}\BibitemShut {NoStop}%
\bibitem [{\citenamefont {Bart\'ok}\ \emph {et~al.}(2010)\citenamefont {Bart\'ok}, \citenamefont {Payne}, \citenamefont {Kondor},\ and\ \citenamefont {Cs\'anyi}}]{Bartok2010}%
  \BibitemOpen
  \bibfield  {author} {\bibinfo {author} {\bibfnamefont {A.~P.}\ \bibnamefont {Bart\'ok}}, \bibinfo {author} {\bibfnamefont {M.~C.}\ \bibnamefont {Payne}}, \bibinfo {author} {\bibfnamefont {R.}~\bibnamefont {Kondor}},\ and\ \bibinfo {author} {\bibfnamefont {G.}~\bibnamefont {Cs\'anyi}},\ }\href {https://doi.org/10.1103/PhysRevLett.104.136403} {\bibfield  {journal} {\bibinfo  {journal} {Phys. Rev. Lett.}\ }\textbf {\bibinfo {volume} {104}},\ \bibinfo {pages} {136403} (\bibinfo {year} {2010})}\BibitemShut {NoStop}%
\bibitem [{\citenamefont {Drautz}(2019)}]{Drautz2019}%
  \BibitemOpen
  \bibfield  {author} {\bibinfo {author} {\bibfnamefont {R.}~\bibnamefont {Drautz}},\ }\href {https://doi.org/10.1103/PhysRevB.99.014104} {\bibfield  {journal} {\bibinfo  {journal} {Phys. Rev. B}\ }\textbf {\bibinfo {volume} {99}},\ \bibinfo {pages} {014104} (\bibinfo {year} {2019})}\BibitemShut {NoStop}%
\bibitem [{\citenamefont {Daw}\ and\ \citenamefont {Baskes}(1984)}]{Daw1984}%
  \BibitemOpen
  \bibfield  {author} {\bibinfo {author} {\bibfnamefont {M.~S.}\ \bibnamefont {Daw}}\ and\ \bibinfo {author} {\bibfnamefont {M.~I.}\ \bibnamefont {Baskes}},\ }\href {https://doi.org/10.1103/PhysRevB.29.6443} {\bibfield  {journal} {\bibinfo  {journal} {Phys. Rev. B}\ }\textbf {\bibinfo {volume} {29}},\ \bibinfo {pages} {6443} (\bibinfo {year} {1984})}\BibitemShut {NoStop}%
\bibitem [{\citenamefont {Mishin}\ \emph {et~al.}(2005)\citenamefont {Mishin}, \citenamefont {Mehl},\ and\ \citenamefont {Papaconstantopoulos}}]{Mishin2005}%
  \BibitemOpen
  \bibfield  {author} {\bibinfo {author} {\bibfnamefont {Y.}~\bibnamefont {Mishin}}, \bibinfo {author} {\bibfnamefont {M.~J.}\ \bibnamefont {Mehl}},\ and\ \bibinfo {author} {\bibfnamefont {D.~A.}\ \bibnamefont {Papaconstantopoulos}},\ }\href {https://doi.org/10.1016/j.actamat.2005.05.001} {\bibfield  {journal} {\bibinfo  {journal} {Acta Mater.}\ }\textbf {\bibinfo {volume} {53}},\ \bibinfo {pages} {4029} (\bibinfo {year} {2005})}\BibitemShut {NoStop}%
\bibitem [{\citenamefont {Batzner}\ \emph {et~al.}(2022)\citenamefont {Batzner}, \citenamefont {Musaelian}, \citenamefont {Sun}, \citenamefont {Geiger}, \citenamefont {Mailoa}, \citenamefont {Kornbluth}, \citenamefont {Molinari}, \citenamefont {Smidt},\ and\ \citenamefont {Kozinsky}}]{Batzner2022}%
  \BibitemOpen
  \bibfield  {author} {\bibinfo {author} {\bibfnamefont {S.}~\bibnamefont {Batzner}}, \bibinfo {author} {\bibfnamefont {A.}~\bibnamefont {Musaelian}}, \bibinfo {author} {\bibfnamefont {L.}~\bibnamefont {Sun}}, \bibinfo {author} {\bibfnamefont {M.}~\bibnamefont {Geiger}}, \bibinfo {author} {\bibfnamefont {J.~P.}\ \bibnamefont {Mailoa}}, \bibinfo {author} {\bibfnamefont {M.}~\bibnamefont {Kornbluth}}, \bibinfo {author} {\bibfnamefont {N.}~\bibnamefont {Molinari}}, \bibinfo {author} {\bibfnamefont {T.~E.}\ \bibnamefont {Smidt}},\ and\ \bibinfo {author} {\bibfnamefont {B.}~\bibnamefont {Kozinsky}},\ }\href {https://doi.org/10.1038/s41467-022-29939-5} {\bibfield  {journal} {\bibinfo  {journal} {Nat. Commun.}\ }\textbf {\bibinfo {volume} {13}},\ \bibinfo {pages} {2453} (\bibinfo {year} {2022})}\BibitemShut {NoStop}%
\bibitem [{\citenamefont {Kavanagh}\ \emph {et~al.}(2026)\citenamefont {Kavanagh}, \citenamefont {Tan}, \citenamefont {Wang}, \citenamefont {Descoteaux}, \citenamefont {de~Miranda~Nascimento}, \citenamefont {Unneberg}, \citenamefont {Zichi}, \citenamefont {Libbi}, \citenamefont {Rivano}, \citenamefont {Glover}, \citenamefont {Bharadwaj}, \citenamefont {Johansson}, \citenamefont {Witt}, \citenamefont {Musaelian},\ and\ \citenamefont {Kozinsky}}]{Kavanagh2026}%
  \BibitemOpen
  \bibfield  {author} {\bibinfo {author} {\bibfnamefont {S.~R.}\ \bibnamefont {Kavanagh}}, \bibinfo {author} {\bibfnamefont {C.~W.}\ \bibnamefont {Tan}}, \bibinfo {author} {\bibfnamefont {M.}~\bibnamefont {Wang}}, \bibinfo {author} {\bibfnamefont {M.~L.}\ \bibnamefont {Descoteaux}}, \bibinfo {author} {\bibfnamefont {G.}~\bibnamefont {de~Miranda~Nascimento}}, \bibinfo {author} {\bibfnamefont {U.}~\bibnamefont {Unneberg}}, \bibinfo {author} {\bibfnamefont {L.}~\bibnamefont {Zichi}}, \bibinfo {author} {\bibfnamefont {F.}~\bibnamefont {Libbi}}, \bibinfo {author} {\bibfnamefont {N.}~\bibnamefont {Rivano}}, \bibinfo {author} {\bibfnamefont {A.}~\bibnamefont {Glover}}, \bibinfo {author} {\bibfnamefont {V.}~\bibnamefont {Bharadwaj}}, \bibinfo {author} {\bibfnamefont {A.}~\bibnamefont {Johansson}}, \bibinfo {author} {\bibfnamefont {W.~C.}\ \bibnamefont {Witt}}, \bibinfo {author} {\bibfnamefont {A.}~\bibnamefont {Musaelian}},\ and\ \bibinfo {author} {\bibfnamefont {B.}~\bibnamefont {Kozinsky}},\ }\href@noop {}
  {\bibinfo {title} {Fast and accurate equivariant foundation models for atomistic simulation}} (\bibinfo {year} {2026}),\ \Eprint {https://arxiv.org/abs/2607.28461} {arXiv:2607.28461} \BibitemShut {NoStop}%
\bibitem [{\citenamefont {Chen}\ and\ \citenamefont {Ong}(2022)}]{Chen2022}%
  \BibitemOpen
  \bibfield  {author} {\bibinfo {author} {\bibfnamefont {C.}~\bibnamefont {Chen}}\ and\ \bibinfo {author} {\bibfnamefont {S.~P.}\ \bibnamefont {Ong}},\ }\href {https://doi.org/10.1038/s43588-022-00349-3} {\bibfield  {journal} {\bibinfo  {journal} {Nat. Comput. Sci.}\ }\textbf {\bibinfo {volume} {2}},\ \bibinfo {pages} {718} (\bibinfo {year} {2022})}\BibitemShut {NoStop}%
\bibitem [{\citenamefont {Batatia}\ \emph {et~al.}(2022)\citenamefont {Batatia}, \citenamefont {Kov\'acs}, \citenamefont {Simm}, \citenamefont {Ortner},\ and\ \citenamefont {Cs\'anyi}}]{Batatia2022}%
  \BibitemOpen
  \bibfield  {author} {\bibinfo {author} {\bibfnamefont {I.}~\bibnamefont {Batatia}}, \bibinfo {author} {\bibfnamefont {D.~P.}\ \bibnamefont {Kov\'acs}}, \bibinfo {author} {\bibfnamefont {G.~N.~C.}\ \bibnamefont {Simm}}, \bibinfo {author} {\bibfnamefont {C.}~\bibnamefont {Ortner}},\ and\ \bibinfo {author} {\bibfnamefont {G.}~\bibnamefont {Cs\'anyi}},\ }in\ \href@noop {} {\emph {\bibinfo {booktitle} {Advances in Neural Information Processing Systems}}},\ Vol.~\bibinfo {volume} {35}\ (\bibinfo {year} {2022})\ p.\ \bibinfo {pages} {11423}\BibitemShut {NoStop}%
\bibitem [{\citenamefont {Batatia}\ \emph {et~al.}(2025)\citenamefont {Batatia} \emph {et~al.}}]{Batatia2025}%
  \BibitemOpen
  \bibfield  {author} {\bibinfo {author} {\bibfnamefont {I.}~\bibnamefont {Batatia}} \emph {et~al.},\ }\href {https://doi.org/10.1063/5.0297006} {\bibfield  {journal} {\bibinfo  {journal} {J. Chem. Phys.}\ }\textbf {\bibinfo {volume} {163}},\ \bibinfo {pages} {184110} (\bibinfo {year} {2025})}\BibitemShut {NoStop}%
\bibitem [{\citenamefont {Pozdnyakov}\ and\ \citenamefont {Ceriotti}(2023)}]{Pozdnyakov2023}%
  \BibitemOpen
  \bibfield  {author} {\bibinfo {author} {\bibfnamefont {S.~N.}\ \bibnamefont {Pozdnyakov}}\ and\ \bibinfo {author} {\bibfnamefont {M.}~\bibnamefont {Ceriotti}},\ }in\ \href@noop {} {\emph {\bibinfo {booktitle} {Advances in Neural Information Processing Systems}}},\ Vol.~\bibinfo {volume} {36}\ (\bibinfo {year} {2023})\ p.\ \bibinfo {pages} {79469}\BibitemShut {NoStop}%
\bibitem [{\citenamefont {Mazitov}\ \emph {et~al.}(2025)\citenamefont {Mazitov}, \citenamefont {Bigi}, \citenamefont {Kellner}, \citenamefont {Pegolo}, \citenamefont {Tisi}, \citenamefont {Fraux}, \citenamefont {Pozdnyakov}, \citenamefont {Loche},\ and\ \citenamefont {Ceriotti}}]{Mazitov2025}%
  \BibitemOpen
  \bibfield  {author} {\bibinfo {author} {\bibfnamefont {A.}~\bibnamefont {Mazitov}}, \bibinfo {author} {\bibfnamefont {F.}~\bibnamefont {Bigi}}, \bibinfo {author} {\bibfnamefont {M.}~\bibnamefont {Kellner}}, \bibinfo {author} {\bibfnamefont {P.}~\bibnamefont {Pegolo}}, \bibinfo {author} {\bibfnamefont {D.}~\bibnamefont {Tisi}}, \bibinfo {author} {\bibfnamefont {G.}~\bibnamefont {Fraux}}, \bibinfo {author} {\bibfnamefont {S.}~\bibnamefont {Pozdnyakov}}, \bibinfo {author} {\bibfnamefont {P.}~\bibnamefont {Loche}},\ and\ \bibinfo {author} {\bibfnamefont {M.}~\bibnamefont {Ceriotti}},\ }\href {https://doi.org/10.1038/s41467-025-65662-7} {\bibfield  {journal} {\bibinfo  {journal} {Nat. Commun.}\ }\textbf {\bibinfo {volume} {16}},\ \bibinfo {pages} {10653} (\bibinfo {year} {2025})}\BibitemShut {NoStop}%
\bibitem [{\citenamefont {Bochkarev}\ \emph {et~al.}(2024)\citenamefont {Bochkarev}, \citenamefont {Lysogorskiy},\ and\ \citenamefont {Drautz}}]{Bochkarev2024}%
  \BibitemOpen
  \bibfield  {author} {\bibinfo {author} {\bibfnamefont {A.}~\bibnamefont {Bochkarev}}, \bibinfo {author} {\bibfnamefont {Y.}~\bibnamefont {Lysogorskiy}},\ and\ \bibinfo {author} {\bibfnamefont {R.}~\bibnamefont {Drautz}},\ }\href {https://doi.org/10.1103/PhysRevX.14.021036} {\bibfield  {journal} {\bibinfo  {journal} {Phys. Rev. X}\ }\textbf {\bibinfo {volume} {14}},\ \bibinfo {pages} {021036} (\bibinfo {year} {2024})}\BibitemShut {NoStop}%
\bibitem [{\citenamefont {Lysogorskiy}\ \emph {et~al.}(2026)\citenamefont {Lysogorskiy}, \citenamefont {Bochkarev},\ and\ \citenamefont {Drautz}}]{Lysogorskiy2026}%
  \BibitemOpen
  \bibfield  {author} {\bibinfo {author} {\bibfnamefont {Y.}~\bibnamefont {Lysogorskiy}}, \bibinfo {author} {\bibfnamefont {A.}~\bibnamefont {Bochkarev}},\ and\ \bibinfo {author} {\bibfnamefont {R.}~\bibnamefont {Drautz}},\ }\href {https://doi.org/10.1038/s41524-026-01979-1} {\bibfield  {journal} {\bibinfo  {journal} {npj Comput. Mater.}\ }\textbf {\bibinfo {volume} {12}},\ \bibinfo {pages} {114} (\bibinfo {year} {2026})}\BibitemShut {NoStop}%
\bibitem [{\citenamefont {Deng}\ \emph {et~al.}(2023)\citenamefont {Deng}, \citenamefont {Zhong}, \citenamefont {Jun}, \citenamefont {Riebesell}, \citenamefont {Han}, \citenamefont {Bartel},\ and\ \citenamefont {Ceder}}]{Deng2023}%
  \BibitemOpen
  \bibfield  {author} {\bibinfo {author} {\bibfnamefont {B.}~\bibnamefont {Deng}}, \bibinfo {author} {\bibfnamefont {P.}~\bibnamefont {Zhong}}, \bibinfo {author} {\bibfnamefont {K.}~\bibnamefont {Jun}}, \bibinfo {author} {\bibfnamefont {J.}~\bibnamefont {Riebesell}}, \bibinfo {author} {\bibfnamefont {K.}~\bibnamefont {Han}}, \bibinfo {author} {\bibfnamefont {C.~J.}\ \bibnamefont {Bartel}},\ and\ \bibinfo {author} {\bibfnamefont {G.}~\bibnamefont {Ceder}},\ }\href {https://doi.org/10.1038/s42256-023-00716-3} {\bibfield  {journal} {\bibinfo  {journal} {Nat. Mach. Intell.}\ }\textbf {\bibinfo {volume} {5}},\ \bibinfo {pages} {1031} (\bibinfo {year} {2023})}\BibitemShut {NoStop}%
\bibitem [{\citenamefont {Barros-Luque}\ \emph {et~al.}(2026)\citenamefont {Barros-Luque}, \citenamefont {Shuaibi}, \citenamefont {Fu}, \citenamefont {Wood}, \citenamefont {Dzamba}, \citenamefont {Gao}, \citenamefont {Rizvi}, \citenamefont {Zitnick},\ and\ \citenamefont {Ulissi}}]{BarrosLuque2026}%
  \BibitemOpen
  \bibfield  {author} {\bibinfo {author} {\bibfnamefont {L.}~\bibnamefont {Barros-Luque}}, \bibinfo {author} {\bibfnamefont {M.}~\bibnamefont {Shuaibi}}, \bibinfo {author} {\bibfnamefont {X.}~\bibnamefont {Fu}}, \bibinfo {author} {\bibfnamefont {B.~M.}\ \bibnamefont {Wood}}, \bibinfo {author} {\bibfnamefont {M.}~\bibnamefont {Dzamba}}, \bibinfo {author} {\bibfnamefont {M.}~\bibnamefont {Gao}}, \bibinfo {author} {\bibfnamefont {A.}~\bibnamefont {Rizvi}}, \bibinfo {author} {\bibfnamefont {C.~L.}\ \bibnamefont {Zitnick}},\ and\ \bibinfo {author} {\bibfnamefont {Z.~W.}\ \bibnamefont {Ulissi}},\ }\href {https://doi.org/10.1038/s43588-026-00996-w} {\bibfield  {journal} {\bibinfo  {journal} {Nat. Comput. Sci.}\ }\textbf {\bibinfo {volume} {6}},\ \bibinfo {pages} {642} (\bibinfo {year} {2026})}\BibitemShut {NoStop}%
\bibitem [{\citenamefont {Riebesell}\ \emph {et~al.}(2025)\citenamefont {Riebesell}, \citenamefont {Goodall}, \citenamefont {Benner}, \citenamefont {Chiang}, \citenamefont {Deng}, \citenamefont {Ceder}, \citenamefont {Asta}, \citenamefont {Lee}, \citenamefont {Jain},\ and\ \citenamefont {Persson}}]{Riebesell2025}%
  \BibitemOpen
  \bibfield  {author} {\bibinfo {author} {\bibfnamefont {J.}~\bibnamefont {Riebesell}}, \bibinfo {author} {\bibfnamefont {R.~E.~A.}\ \bibnamefont {Goodall}}, \bibinfo {author} {\bibfnamefont {P.}~\bibnamefont {Benner}}, \bibinfo {author} {\bibfnamefont {Y.}~\bibnamefont {Chiang}}, \bibinfo {author} {\bibfnamefont {B.}~\bibnamefont {Deng}}, \bibinfo {author} {\bibfnamefont {G.}~\bibnamefont {Ceder}}, \bibinfo {author} {\bibfnamefont {M.}~\bibnamefont {Asta}}, \bibinfo {author} {\bibfnamefont {A.~A.}\ \bibnamefont {Lee}}, \bibinfo {author} {\bibfnamefont {A.}~\bibnamefont {Jain}},\ and\ \bibinfo {author} {\bibfnamefont {K.~A.}\ \bibnamefont {Persson}},\ }\href {https://doi.org/10.1038/s42256-025-01055-1} {\bibfield  {journal} {\bibinfo  {journal} {Nat. Mach. Intell.}\ }\textbf {\bibinfo {volume} {7}},\ \bibinfo {pages} {836} (\bibinfo {year} {2025})}\BibitemShut {NoStop}%
\bibitem [{\citenamefont {Loew}\ \emph {et~al.}(2025)\citenamefont {Loew}, \citenamefont {Sun}, \citenamefont {Wang}, \citenamefont {Botti},\ and\ \citenamefont {Marques}}]{Loew2025}%
  \BibitemOpen
  \bibfield  {author} {\bibinfo {author} {\bibfnamefont {A.}~\bibnamefont {Loew}}, \bibinfo {author} {\bibfnamefont {D.}~\bibnamefont {Sun}}, \bibinfo {author} {\bibfnamefont {H.-C.}\ \bibnamefont {Wang}}, \bibinfo {author} {\bibfnamefont {S.}~\bibnamefont {Botti}},\ and\ \bibinfo {author} {\bibfnamefont {M.~A.~L.}\ \bibnamefont {Marques}},\ }\href {https://doi.org/10.1038/s41524-025-01650-1} {\bibfield  {journal} {\bibinfo  {journal} {npj Comput. Mater.}\ }\textbf {\bibinfo {volume} {11}},\ \bibinfo {pages} {178} (\bibinfo {year} {2025})}\BibitemShut {NoStop}%
\bibitem [{\citenamefont {P{\'o}ta}\ \emph {et~al.}(2026)\citenamefont {P{\'o}ta}, \citenamefont {Ahlawat}, \citenamefont {Cs{\'a}nyi},\ and\ \citenamefont {Simoncelli}}]{Pota2026}%
  \BibitemOpen
  \bibfield  {author} {\bibinfo {author} {\bibfnamefont {B.}~\bibnamefont {P{\'o}ta}}, \bibinfo {author} {\bibfnamefont {P.}~\bibnamefont {Ahlawat}}, \bibinfo {author} {\bibfnamefont {G.}~\bibnamefont {Cs{\'a}nyi}},\ and\ \bibinfo {author} {\bibfnamefont {M.}~\bibnamefont {Simoncelli}},\ }\href {https://doi.org/10.1038/s41467-026-76391-w} {\bibfield  {journal} {\bibinfo  {journal} {Nat. Commun.}\ }\textbf {\bibinfo {volume} {17}},\ \bibinfo {pages} {10566} (\bibinfo {year} {2026})}\BibitemShut {NoStop}%
\bibitem [{\citenamefont {Swinburne}\ \emph {et~al.}(2020)\citenamefont {Swinburne}, \citenamefont {Janssen}, \citenamefont {Todorova}, \citenamefont {Simpson}, \citenamefont {Plechac}, \citenamefont {Luskin},\ and\ \citenamefont {Neugebauer}}]{Swinburne2020}%
  \BibitemOpen
  \bibfield  {author} {\bibinfo {author} {\bibfnamefont {T.~D.}\ \bibnamefont {Swinburne}}, \bibinfo {author} {\bibfnamefont {J.}~\bibnamefont {Janssen}}, \bibinfo {author} {\bibfnamefont {M.}~\bibnamefont {Todorova}}, \bibinfo {author} {\bibfnamefont {G.}~\bibnamefont {Simpson}}, \bibinfo {author} {\bibfnamefont {P.}~\bibnamefont {Plechac}}, \bibinfo {author} {\bibfnamefont {M.}~\bibnamefont {Luskin}},\ and\ \bibinfo {author} {\bibfnamefont {J.}~\bibnamefont {Neugebauer}},\ }\href {https://doi.org/10.1103/PhysRevB.102.100101} {\bibfield  {journal} {\bibinfo  {journal} {Phys. Rev. B}\ }\textbf {\bibinfo {volume} {102}},\ \bibinfo {pages} {100101(R)} (\bibinfo {year} {2020})}\BibitemShut {NoStop}%
\bibitem [{\citenamefont {Dsouza}\ \emph {et~al.}(2024)\citenamefont {Dsouza}, \citenamefont {Poul}, \citenamefont {Huber}, \citenamefont {Swinburne},\ and\ \citenamefont {Neugebauer}}]{Dsouza2024}%
  \BibitemOpen
  \bibfield  {author} {\bibinfo {author} {\bibfnamefont {R.}~\bibnamefont {Dsouza}}, \bibinfo {author} {\bibfnamefont {M.}~\bibnamefont {Poul}}, \bibinfo {author} {\bibfnamefont {L.}~\bibnamefont {Huber}}, \bibinfo {author} {\bibfnamefont {T.~D.}\ \bibnamefont {Swinburne}},\ and\ \bibinfo {author} {\bibfnamefont {J.}~\bibnamefont {Neugebauer}},\ }\href {https://doi.org/10.1103/PhysRevB.109.064108} {\bibfield  {journal} {\bibinfo  {journal} {Phys. Rev. B}\ }\textbf {\bibinfo {volume} {109}},\ \bibinfo {pages} {064108} (\bibinfo {year} {2024})}\BibitemShut {NoStop}%
\bibitem [{\citenamefont {Dsouza}(2025)}]{Dsouza2025thesis}%
  \BibitemOpen
  \bibfield  {author} {\bibinfo {author} {\bibfnamefont {R.}~\bibnamefont {Dsouza}},\ }\emph {\bibinfo {title} {Surrogate models for sampling-free thermodynamics of crystalline materials}},\ \href {https://doi.org/10.13154/294-13330} {Ph.D. thesis},\ \bibinfo  {school} {Ruhr-Universit{\"a}t Bochum}, \bibinfo {address} {Bochum, Germany} (\bibinfo {year} {2025})\BibitemShut {NoStop}%
\bibitem [{\citenamefont {Togo}\ \emph {et~al.}(2024)\citenamefont {Togo}, \citenamefont {Shinohara},\ and\ \citenamefont {Tanaka}}]{Togo2024}%
  \BibitemOpen
  \bibfield  {author} {\bibinfo {author} {\bibfnamefont {A.}~\bibnamefont {Togo}}, \bibinfo {author} {\bibfnamefont {K.}~\bibnamefont {Shinohara}},\ and\ \bibinfo {author} {\bibfnamefont {I.}~\bibnamefont {Tanaka}},\ }\href {https://doi.org/10.1080/27660400.2024.2384822} {\bibfield  {journal} {\bibinfo  {journal} {Sci. Technol. Adv. Mater. Methods}\ }\textbf {\bibinfo {volume} {4}},\ \bibinfo {pages} {2384822} (\bibinfo {year} {2024})}\BibitemShut {NoStop}%
\bibitem [{\citenamefont {Anderson}(2003)}]{Anderson2003}%
  \BibitemOpen
  \bibfield  {author} {\bibinfo {author} {\bibfnamefont {T.~W.}\ \bibnamefont {Anderson}},\ }\href@noop {} {\emph {\bibinfo {title} {An Introduction to Multivariate Statistical Analysis}}},\ \bibinfo {edition} {3rd}\ ed.\ (\bibinfo  {publisher} {Wiley-Interscience},\ \bibinfo {address} {Hoboken, NJ},\ \bibinfo {year} {2003})\BibitemShut {NoStop}%
\bibitem [{\citenamefont {Bishop}(2006)}]{Bishop2006}%
  \BibitemOpen
  \bibfield  {author} {\bibinfo {author} {\bibfnamefont {C.~M.}\ \bibnamefont {Bishop}},\ }\href@noop {} {\emph {\bibinfo {title} {Pattern Recognition and Machine Learning}}},\ Information Science and Statistics\ (\bibinfo  {publisher} {Springer},\ \bibinfo {address} {New York},\ \bibinfo {year} {2006})\BibitemShut {NoStop}%
\bibitem [{\citenamefont {Penrose}(1955)}]{Penrose1955}%
  \BibitemOpen
  \bibfield  {author} {\bibinfo {author} {\bibfnamefont {R.}~\bibnamefont {Penrose}},\ }\href {https://doi.org/10.1017/S0305004100030401} {\bibfield  {journal} {\bibinfo  {journal} {Math. Proc. Cambridge Philos. Soc.}\ }\textbf {\bibinfo {volume} {51}},\ \bibinfo {pages} {406} (\bibinfo {year} {1955})}\BibitemShut {NoStop}%
\bibitem [{\citenamefont {Graf}\ \emph {et~al.}(2011)\citenamefont {Graf}, \citenamefont {Felser},\ and\ \citenamefont {Parkin}}]{Graf2011}%
  \BibitemOpen
  \bibfield  {author} {\bibinfo {author} {\bibfnamefont {T.}~\bibnamefont {Graf}}, \bibinfo {author} {\bibfnamefont {C.}~\bibnamefont {Felser}},\ and\ \bibinfo {author} {\bibfnamefont {S.~S.~P.}\ \bibnamefont {Parkin}},\ }\href {https://doi.org/10.1016/j.progsolidstchem.2011.02.001} {\bibfield  {journal} {\bibinfo  {journal} {Prog. Solid State Chem.}\ }\textbf {\bibinfo {volume} {39}},\ \bibinfo {pages} {1} (\bibinfo {year} {2011})}\BibitemShut {NoStop}%
\bibitem [{\citenamefont {Mendelev}\ \emph {et~al.}(2008)\citenamefont {Mendelev}, \citenamefont {Kramer}, \citenamefont {Becker},\ and\ \citenamefont {Asta}}]{Mendelev2008}%
  \BibitemOpen
  \bibfield  {author} {\bibinfo {author} {\bibfnamefont {M.~I.}\ \bibnamefont {Mendelev}}, \bibinfo {author} {\bibfnamefont {M.~J.}\ \bibnamefont {Kramer}}, \bibinfo {author} {\bibfnamefont {C.~A.}\ \bibnamefont {Becker}},\ and\ \bibinfo {author} {\bibfnamefont {M.}~\bibnamefont {Asta}},\ }\href {https://doi.org/10.1080/14786430802206482} {\bibfield  {journal} {\bibinfo  {journal} {Philos. Mag.}\ }\textbf {\bibinfo {volume} {88}},\ \bibinfo {pages} {1723} (\bibinfo {year} {2008})}\BibitemShut {NoStop}%
\bibitem [{\citenamefont {Starikov}\ \emph {et~al.}(2021)\citenamefont {Starikov}, \citenamefont {Smirnova}, \citenamefont {Pradhan}, \citenamefont {Lysogorskiy}, \citenamefont {Chapman}, \citenamefont {Mrovec},\ and\ \citenamefont {Drautz}}]{Starikov2021}%
  \BibitemOpen
  \bibfield  {author} {\bibinfo {author} {\bibfnamefont {S.}~\bibnamefont {Starikov}}, \bibinfo {author} {\bibfnamefont {D.}~\bibnamefont {Smirnova}}, \bibinfo {author} {\bibfnamefont {T.}~\bibnamefont {Pradhan}}, \bibinfo {author} {\bibfnamefont {Y.}~\bibnamefont {Lysogorskiy}}, \bibinfo {author} {\bibfnamefont {H.}~\bibnamefont {Chapman}}, \bibinfo {author} {\bibfnamefont {M.}~\bibnamefont {Mrovec}},\ and\ \bibinfo {author} {\bibfnamefont {R.}~\bibnamefont {Drautz}},\ }\href {https://doi.org/10.1103/PhysRevMaterials.5.063607} {\bibfield  {journal} {\bibinfo  {journal} {Phys. Rev. Materials}\ }\textbf {\bibinfo {volume} {5}},\ \bibinfo {pages} {063607} (\bibinfo {year} {2021})}\BibitemShut {NoStop}%
\bibitem [{\citenamefont {Sun}\ \emph {et~al.}(2006)\citenamefont {Sun}, \citenamefont {Mendelev}, \citenamefont {Becker}, \citenamefont {Kudin}, \citenamefont {Haxhimali}, \citenamefont {Asta}, \citenamefont {Hoyt}, \citenamefont {Karma},\ and\ \citenamefont {Srolovitz}}]{Sun2006}%
  \BibitemOpen
  \bibfield  {author} {\bibinfo {author} {\bibfnamefont {D.~Y.}\ \bibnamefont {Sun}}, \bibinfo {author} {\bibfnamefont {M.~I.}\ \bibnamefont {Mendelev}}, \bibinfo {author} {\bibfnamefont {C.~A.}\ \bibnamefont {Becker}}, \bibinfo {author} {\bibfnamefont {K.}~\bibnamefont {Kudin}}, \bibinfo {author} {\bibfnamefont {T.}~\bibnamefont {Haxhimali}}, \bibinfo {author} {\bibfnamefont {M.}~\bibnamefont {Asta}}, \bibinfo {author} {\bibfnamefont {J.~J.}\ \bibnamefont {Hoyt}}, \bibinfo {author} {\bibfnamefont {A.}~\bibnamefont {Karma}},\ and\ \bibinfo {author} {\bibfnamefont {D.~J.}\ \bibnamefont {Srolovitz}},\ }\href {https://doi.org/10.1103/PhysRevB.73.024116} {\bibfield  {journal} {\bibinfo  {journal} {Phys. Rev. B}\ }\textbf {\bibinfo {volume} {73}},\ \bibinfo {pages} {024116} (\bibinfo {year} {2006})}\BibitemShut {NoStop}%
\bibitem [{\citenamefont {Mendelev}\ \emph {et~al.}(2005)\citenamefont {Mendelev}, \citenamefont {Srolovitz}, \citenamefont {Ackland},\ and\ \citenamefont {Han}}]{Mendelev2005}%
  \BibitemOpen
  \bibfield  {author} {\bibinfo {author} {\bibfnamefont {M.~I.}\ \bibnamefont {Mendelev}}, \bibinfo {author} {\bibfnamefont {D.~J.}\ \bibnamefont {Srolovitz}}, \bibinfo {author} {\bibfnamefont {G.~J.}\ \bibnamefont {Ackland}},\ and\ \bibinfo {author} {\bibfnamefont {S.}~\bibnamefont {Han}},\ }\href {https://doi.org/10.1557/JMR.2005.0024} {\bibfield  {journal} {\bibinfo  {journal} {J. Mater. Res.}\ }\textbf {\bibinfo {volume} {20}},\ \bibinfo {pages} {208} (\bibinfo {year} {2005})}\BibitemShut {NoStop}%
\bibitem [{\citenamefont {Hayakawa}\ and\ \citenamefont {Xu}(2024)}]{Hayakawa2024}%
  \BibitemOpen
  \bibfield  {author} {\bibinfo {author} {\bibfnamefont {S.}~\bibnamefont {Hayakawa}}\ and\ \bibinfo {author} {\bibfnamefont {H.}~\bibnamefont {Xu}},\ }\href {https://doi.org/10.1016/j.commatsci.2023.112614} {\bibfield  {journal} {\bibinfo  {journal} {Comput. Mater. Sci.}\ }\textbf {\bibinfo {volume} {232}},\ \bibinfo {pages} {112614} (\bibinfo {year} {2024})}\BibitemShut {NoStop}%
\bibitem [{\citenamefont {Dsouza}\ \emph {et~al.}(2026)\citenamefont {Dsouza}, \citenamefont {Swinburne},\ and\ \citenamefont {Neugebauer}}]{Dsouza2026data}%
  \BibitemOpen
  \bibfield  {author} {\bibinfo {author} {\bibfnamefont {R.}~\bibnamefont {Dsouza}}, \bibinfo {author} {\bibfnamefont {T.~D.}\ \bibnamefont {Swinburne}},\ and\ \bibinfo {author} {\bibfnamefont {J.}~\bibnamefont {Neugebauer}},\ }\href {https://doi.org/10.5281/zenodo.24466123} {\bibinfo {title} {Data for: {T}hermodynamics of complex crystals from a generalized mean-field bond model, version 1.0.0}},\ \bibinfo {howpublished} {Zenodo} (\bibinfo {year} {2026})\BibitemShut {NoStop}%
\bibitem [{\citenamefont {Thompson}\ \emph {et~al.}(2022)\citenamefont {Thompson}, \citenamefont {Aktulga}, \citenamefont {Berger}, \citenamefont {Bolintineanu}, \citenamefont {Brown}, \citenamefont {Crozier}, \citenamefont {in~'t Veld}, \citenamefont {Kohlmeyer}, \citenamefont {Moore}, \citenamefont {Nguyen}, \citenamefont {Shan}, \citenamefont {Stevens}, \citenamefont {Tranchida}, \citenamefont {Trott},\ and\ \citenamefont {Plimpton}}]{Thompson2022}%
  \BibitemOpen
  \bibfield  {author} {\bibinfo {author} {\bibfnamefont {A.~P.}\ \bibnamefont {Thompson}}, \bibinfo {author} {\bibfnamefont {H.~M.}\ \bibnamefont {Aktulga}}, \bibinfo {author} {\bibfnamefont {R.}~\bibnamefont {Berger}}, \bibinfo {author} {\bibfnamefont {D.~S.}\ \bibnamefont {Bolintineanu}}, \bibinfo {author} {\bibfnamefont {W.~M.}\ \bibnamefont {Brown}}, \bibinfo {author} {\bibfnamefont {P.~S.}\ \bibnamefont {Crozier}}, \bibinfo {author} {\bibfnamefont {P.~J.}\ \bibnamefont {in~'t Veld}}, \bibinfo {author} {\bibfnamefont {A.}~\bibnamefont {Kohlmeyer}}, \bibinfo {author} {\bibfnamefont {S.~G.}\ \bibnamefont {Moore}}, \bibinfo {author} {\bibfnamefont {T.~D.}\ \bibnamefont {Nguyen}}, \bibinfo {author} {\bibfnamefont {R.}~\bibnamefont {Shan}}, \bibinfo {author} {\bibfnamefont {M.~J.}\ \bibnamefont {Stevens}}, \bibinfo {author} {\bibfnamefont {J.}~\bibnamefont {Tranchida}}, \bibinfo {author} {\bibfnamefont {C.}~\bibnamefont {Trott}},\ and\ \bibinfo {author} {\bibfnamefont {S.~J.}\ \bibnamefont {Plimpton}},\ }\href
  {https://doi.org/10.1016/j.cpc.2021.108171} {\bibfield  {journal} {\bibinfo  {journal} {Comput. Phys. Commun.}\ }\textbf {\bibinfo {volume} {271}},\ \bibinfo {pages} {108171} (\bibinfo {year} {2022})}\BibitemShut {NoStop}%
\bibitem [{\citenamefont {Dsouza}(2026)}]{Dsouza2026code}%
  \BibitemOpen
  \bibfield  {author} {\bibinfo {author} {\bibfnamefont {R.}~\bibnamefont {Dsouza}},\ }\href {https://doi.org/10.5281/zenodo.24493341} {\bibinfo {title} {mean-field-bond-model, version 1.0.0}},\ \bibinfo {howpublished} {Zenodo} (\bibinfo {year} {2026}),\ \bibinfo {note} {source code at \url{https://github.com/raynol-dsouza/mean-field-bond-model}}\BibitemShut {NoStop}%
\bibitem [{\citenamefont {Herrero}\ and\ \citenamefont {Ram\'irez}(2014)}]{Herrero2014}%
  \BibitemOpen
  \bibfield  {author} {\bibinfo {author} {\bibfnamefont {C.~P.}\ \bibnamefont {Herrero}}\ and\ \bibinfo {author} {\bibfnamefont {R.}~\bibnamefont {Ram\'irez}},\ }\href {https://doi.org/10.1088/0953-8984/26/23/233201} {\bibfield  {journal} {\bibinfo  {journal} {J. Phys.: Condens. Matter}\ }\textbf {\bibinfo {volume} {26}},\ \bibinfo {pages} {233201} (\bibinfo {year} {2014})}\BibitemShut {NoStop}%
\bibitem [{\citenamefont {Markland}\ and\ \citenamefont {Ceriotti}(2018)}]{Markland2018}%
  \BibitemOpen
  \bibfield  {author} {\bibinfo {author} {\bibfnamefont {T.~E.}\ \bibnamefont {Markland}}\ and\ \bibinfo {author} {\bibfnamefont {M.}~\bibnamefont {Ceriotti}},\ }\href {https://doi.org/10.1038/s41570-017-0109} {\bibfield  {journal} {\bibinfo  {journal} {Nat. Rev. Chem.}\ }\textbf {\bibinfo {volume} {2}},\ \bibinfo {pages} {0109} (\bibinfo {year} {2018})}\BibitemShut {NoStop}%
\bibitem [{\citenamefont {Dsouza}\ \emph {et~al.}(2022)\citenamefont {Dsouza}, \citenamefont {Huber}, \citenamefont {Grabowski},\ and\ \citenamefont {Neugebauer}}]{Dsouza2022}%
  \BibitemOpen
  \bibfield  {author} {\bibinfo {author} {\bibfnamefont {R.}~\bibnamefont {Dsouza}}, \bibinfo {author} {\bibfnamefont {L.}~\bibnamefont {Huber}}, \bibinfo {author} {\bibfnamefont {B.}~\bibnamefont {Grabowski}},\ and\ \bibinfo {author} {\bibfnamefont {J.}~\bibnamefont {Neugebauer}},\ }\href {https://doi.org/10.1103/PhysRevB.105.184111} {\bibfield  {journal} {\bibinfo  {journal} {Phys. Rev. B}\ }\textbf {\bibinfo {volume} {105}},\ \bibinfo {pages} {184111} (\bibinfo {year} {2022})}\BibitemShut {NoStop}%
\end{thebibliography}%

\end{document}